\documentclass[preprint2]{aastex}
\usepackage{graphicx}
\usepackage{longtable,lscape}
 
\shorttitle{Stellar parameters of late-K and M dwarfs}
\shortauthors{Houdebine, Mullan, Paletou \& Gebran}

\begin{document}

\title{The Rotation-Activity Correlations in K and M dwarfs. I. Stellar 
parameters, compilations of $v\sin i$ and $P/\sin i$ for a large sample of 
late-K and M dwarfs \thanks{Based on observations available at 
Observatoire de Haute Provence and the European Southern Observatory 
databases and on Hipparcos parallax measurements.}}

\author{E.R. Houdebine}
\affil{Armagh Observatory, College Hill, BT61 9DG Armagh, Northern Ireland}
\affil{Universit\'e de Toulouse, UPS-Observatoire Midi-Pyr\'en\'ees, IRAP, 
Toulouse, France}
\affil{CNRS, Institut de Recherche en Astrophysique et Plan\'etologie, 14 
av. E. Belin, F--31400 Toulouse, France}
\email{eric\_houdebine@yahoo.fr}

\author{D.J. Mullan}
\affil{Department of Physics and Astronomy, University of Delaware, 
Newark, DE 19716, USA}

\author{F. Paletou}
\affil{Universit\'e de Toulouse, UPS-Observatoire Midi-Pyr\'en\'ees, IRAP, 
Toulouse, France}
\affil{CNRS, Institut de Recherche en Astrophysique et Plan\'etologie, 14 av. 
E. Belin, F--31400 Toulouse, France}

\and 
\author{M. Gebran}
\affil{Department of Physics \& Astronomy, Notre Dame University-Louaize, PO 
Box 72, Zouk Mika\"{e}l, Lebanon}


\begin{abstract}
\small
Reliable determination of Rotation-Activity Correlation (RAC's) depends on 
precise measurements of the following stellar parameters: $T_{eff}$, parallax, 
radius, metallicity, and rotational speed $v\sin i$. In this paper, our goal 
is to focus on the determination of these parameters for a sample of K and M 
dwarfs. In a future paper (Paper II), we will combine our rotational data with 
activity data in order to construct RAC's. 

Here, we report on a determination of effective temperatures 
based on the (R-I)$_C$ color from the calibrations of Mann et al. (2015) 
and Kenyon \& Hartmann (1995) for four samples of late-K, dM2, dM3 and 
dM4 stars. We also determine stellar parameters ($T_{eff}$, $log(g)$ and 
[M/H]) using the PCA-based inversion technique for a sample of 105 late-K 
dwarfs. We compile all effective temperatures from the literature for 
this sample. We determine empirical radius-[M/H] correlations in our 
stellar samples. This allows us to propose new effective temperatures, 
stellar radii, and metallicities for a large sample of 612 late-K and M 
dwarfs. Our mean radii agree well with those of Boyajian et al. (2012).

We analyze HARPS and SOPHIE spectra of 105 late-K dwarfs, and have detected 
$v\sin i$ in 92 stars. In combination with our previous $v\sin i$ 
measurements in M and K dwarfs, we now derive $P/sin i$ measures for a 
sample of 418 K and M dwarfs. We investigate the distributions of $P/sin i$ 
and we show that they are different from one spectral sub-type to another 
at a 99.9\% confidence level.
\end{abstract}
\normalsize


\section{Introduction}

The determination of fundamental stellar parameters ($T_{eff}$, $R_{*}$, [M/H] 
and $v\sin i$), notably for nearby stars, are essential to many astrophysical 
studies. This is especially true for studies of Rotation-Activity Correlations 
(RAC's), i.e., studies of how various indices of magnetic activity (e.g. 
$R'_{HK}$, $L_{X}$) vary as a function of the rotation period of the star. In 
this paper, we focus on quantities which are relevant to determining 
rotational properties, including $T_{eff}$, $R_*$, [M/H], and $v\sin i$. We 
shall postpone the discussion of activity indices to Paper II, when the RAC's 
will be constructed.

Various techniques have been used in the past to derive values of $T_{eff}$ 
and [M/H]. Here we use two different approaches to determine these parameters 
in samples of late-K and M dwarfs. In the first approach, we use the PCA-based 
inversion technique of Paletou et al. (2015a), by applying it to the Na\,{\sc 
i} resonance doublet in our K dwarf sample (105 stars). For this sample of K 
dwarfs, we also compiled all published values of $T_{eff}$ in the literature 
and further derived new values of $T_{eff}$ based on measurements of the 
(R-I)$_{C}$ color: our conversion from (R-I)$_{C}$ to $T_{eff}$ uses the
extensive calibration of Kenyon \& Hartmann (1995). We find good agreement 
between our values rederived from (R-I)$_{C}$ and the published values 
($3\sigma \sim 70$ K).

In the second approach, for our M dwarf samples (M2, M3 and M4), we find that 
a tight correlation between $T_{eff}$ and (R-I)$_{C}$ exists among the recent 
measurements of Mann et al. (2015). It had been shown in the past that the 
(R-I)$_{C}$ color is a good effective temperature diagnostic for M dwarfs (e.g.
 Leggett 1992), with a relatively low sensitivity on the metallicity (e.g. 
Ramirez \& Melendez 2005). In an on-going determination of $T_{eff}$ derived 
from (R-I)${C}$ for samples of M dwarfs ranging in spectral subtype from M2 to 
M7, Houdebine et al. (2016, in preparation) find a very good agreement with a 
compilation of all published values of $T_{eff}$ in the literature. Therefore, 
for our present samples of M2, M3 and M4 dwarfs, we have used the calibration 
of Mann et al. (2015) to determine $T_{eff}$. 

Houdebine (2008, Paper VII) found a relatively tight correlation between [M/H] 
and $R_{*}$ in a sample of M2 dwarfs. [M/H] decreases with decreasing  $R_{*}$ 
in normal dwarfs and subdwarfs, such that all subdwarfs have very low [M/H] 
and $R_{*}$. In the present paper, we perform extensive compilations of [M/H] 
from the literature and correlate [M/H] and $R_{*}$ for our samples of late-K, 
M3, and M4 dwarfs. This allows us to obtain values of [M/H] for all of our 
target stars. The determination of [M/H] will be essential when we eventually 
attempt to construct meaningful RAC's for the Ca\,{\sc ii} line fluxes. The 
reason is that an ongoing NLTE-radiation tranfer modelling study (Houdebine \& 
Panagi 2016, in preparation) of the  Ca\,{\sc ii} line formation as a function 
of the metallicity confirms that the {\em Ca\,{\sc ii} line resonance doublet 
is not a good diagnostic of magnetic activity unless it is first corrected for 
the influence of metallicity}.

As a preliminary to constructing RAC's in K and M dwarfs, we have reported 
over the past several years, improved spectroscopic measurements of rotational 
broadening $v\sin i$ in stars of spectral sub-types dK4 (Houdebine 2011a, 
Paper XVI thereafter), dK6 (present study), dM2 (Houdebine 2008, Paper VIII, 
Houdebine 2010a, Paper XIV), dM3 (Houdebine \& Mullan 2015) and dM4 (Houdebine 
2012a, Paper XVII). An important aspect of our $v\sin i$ measurements has to 
do with the use of the HARPS and SOPHIE spectrographs: these spectrographs are 
designed initially for exo-planet searches and are therefore very stable in 
wavelength in the long term (about 1 m/s for HARPS). These data give us access 
to rotational velocities which are significantly smaller than in previous 
spectroscopic studies of K and M dwarfs. Because of the smallness of some of 
our $v\sin i$ values, we need to verify the precision of our data. We assess 
the uncertainties on our $v\sin i$ measurements in Section~5 below. We shall 
find that the uncertainties $\delta v$ in $v\sin i$ may be as small as  $\pm 
\sigma =0.27\ km ~s^{-1}$ for  $v\sin i<1\ km~s^{-1}$, $\pm \sigma =0.32\ 
km~s^{-1}$ for $1<v\sin i<5\ km~s^{-1}$, and $\pm \sigma=0.60\ km~s^{-1}$ for 
$v\sin i>5\ km~s^{-1}$.

The occurrence of such uncertainties imposes an upper limit $P_{max}$/$\sin i$ 
on the rotational period of stars which are accessible to our approach. E.g., 
for stars near the transition to complete convection (TTCC), where 
$R_{*}\approx$0.3-0.4~$R_{\odot}$, we can in principle measure the rotation 
periods of TTCC stars with confidence only up to values of $P_{max}/\sin i$ = 
2$\pi R_{*}/\delta_{v}\sim$ 15-40 days. With a random distribution of 
inclinations, the average value of $\sin i$ is 2/$\pi$ = 0.64. Therefore, the 
longest rotational periods $P_{max}$ that we can hope to measure near the TTCC 
would be no greater than 10-26 days. For our sample of dK6 stars, where the 
average $R_{*}=0.647 R_{\odot}$ (see Table 6), the upper limit on 
$P_{max}/\sin i$ is 33-66 days. Again correcting for $\sin i$ = 0.64, we find 
$P_{max}$ for dK6 stars to be 21-42 days. Interestingly, these ranges in 
period overlap with the upper limit (about 30 days) which was obtained from 
early studies of rotational modulation in Kepler photometry (Nielsen et al. 
2013). It is encouraging to find that by pushing spectroscopic studies of 
rotation to better limits, we can now study stars whose rotational periods 
would previously have been accessible only to photometric studies. To be sure, 
our approach is still subject to the $\sin i$ limitation, whereas this does 
not affect photometric analysis.

In the present study, we report on a new analysis of high resolution HARPS and 
SOPHIE spectra of 105 late-K dwarfs. We have detected $v\sin i$ in 92 of these 
stars. By combining these data with analyses in our earlier papers, we now 
investigate the distributions of $P/\sin i$ for our samples of stars at 
spectral subtypes dK6, dM2, dM3, and dM4. In total, the present compilation of 
$P/\sin i$ measures includes measurements for 418 different K and M dwarfs. As 
already reported in paper XVI and in Houdebine \& Mullan (2015, hereafter HM), 
we confirm here that these distributions vary with spectral subtype: the 
distributions among dK6, dM2, dM3 and dM4 stars are statistically different 
from one another. These findings imply that in order to study RAC's, one must 
construct RAC's separately for each different spectral sub-type. 

Our data confirm that the mean rotation periods of stars in the range dK4-dM4 
in general decrease with decreasing effective temperature for slow rotators 
(HM). However, we also find that something unusual happens in the rotation 
rates between dM2 and dM4. The overall trend towards decreasing rotation 
period as we go from dK4 to dM4 is interrupted at spectral sub-type dM3 where 
the period increases to a local peak where the rotational period is longer 
than the overall trend would have predicted (HM). When the investigation is 
extended (as in the present paper) to include fast rotators among the dK4-dM4 
stars, one finds the following overall trend: the mean rotation period tends 
to increase slightly from dK4 to dM4. But once again, at dM3, an exception is 
found: the mean rotation period of fast rotators at dM3 is locally 
significantly longer than the overall trend would have predicted (HM). HM 
interpreted the abnormally long rotation periods at sub-type dM3 as possibly 
being associated with the occurrence of increasing coronal loop lengths 
(reported by Mullan et al. 2006). The mean rotation period of the slow 
rotators is an important constraint on the past evolution of the dynamo 
mechanisms and magnetic braking mechanisms (HM and this study). 

The plan of this paper is as follows. In Section 2, we describe how the 
spectroscopic data were selected for a sample of late-K dwarfs. In Section 3, 
we describe how values of $T_{eff}$, $R_*$, and [M/H] have been extracted for 
our target stars, first for the K dwarfs, and then for the M dwarfs. In 
Section~4, we describe how a cross-correlation technique is used to determine 
the value of $v\sin i$ for our sample of dK6 stars using all spectral lines 
within a preferred band of the spectrum. Uncertainties in the values of $v\sin 
i$ for K and M dwarfs are discussed in Section 5. In Section 6, we present the 
distributions of $P/\sin i$ values for the stars of different spectral 
sub-types, and discuss the statistical significance of differences between the 
distributions. Our conclusions are in Section 7. 

We stress that the contents of Paper I are only a first step in a two-step 
process which aims to derive RAC's for each spectral sub-type. The second step 
in this process will be discussed in Paper II. The present paper focusses on 
obtaining reliable data for the ``Rotation" axis of the RAC. The subsequent 
paper will focus at first on obtaining reliable data for the ``Activity" axis 
of the RAC. Once reliable data are available for both axes, a search will then 
be undertaken (in Paper II) to determine if any significant correlation exists 
between Rotation and Activity for each spectral sub-type.

\section{Selection of spectroscopic data}

Stepien (1989, 1993, 1994) has reported that RAC's exhibit certain differences 
at different spectral types. As an extension of this finding, we have found, 
in previous studies (Paper XVIII, HM), that it is important in constructing 
RAC's to select samples of stars with $T_{eff}$ values which are confined 
within a narrow range. There are two principal reasons for this, one related 
to our analysis of chromospheric emission lines, and one related to the choice 
of an optimal set of photospheric absorption lines. 

As regards {\it chromospheric} analysis, it is important to deal with stars 
with closely similar $T_{eff}$ when we are attempting to quantify with as much 
precision as possible the EW of the chromospheric lines (Ca\,{\sc ii} 
resonance doublet and H$_{\alpha}$). These lines inevitably include some 
contributions from the background photospheric continuum and from the 
temperature minimum region (e.g.Houdebine \& Doyle 1994, Cram \& Mullan 1979, 
Houdebine \& Stempels 1997, Paper XV). Initially, our samples of stars had 
been selected for the purpose of chromospheric modelling studies (e.g. 
Houdebine \& Stempels 1997, Houdebine 2009b Paper XII, Houdebine 2010b Paper 
IX), and in these studies the selection of stars with closely similar spectral 
types was essential in order to develop reliable grids of semi-empirical model 
chromospheres, each of which would be superposed on a particular photospheric 
model. 

As regards {\it photospheric} analysis, our principal goal is to extract 
precise $v\sin i$ values by applying cross-correlation methods to multiple 
absorption lines over a $\sim$100~\AA\ swath of spectrum. We have found that 
it is important to select certain optimal ranges of wavelengths so that the 
spectral lines which lie within those ranges will contribute maximally to 
enhancing the precision of $v\sin i$: the precision is optimized in spectral 
ranges where there are large numbers of weak and unsaturated narrow spectral 
lines. As it turns out, the photospheric continuum also contributes to which 
wavelengths ranges are optimal for our purposes: even within the relatively 
narrow range of dK4 to dM4, changes in the continuum from one spectral 
sub-type to another give rise to detectable differences in optimizing the 
choice of the wavelength range. These factors reinforce our decision to study 
homogeneous samples where the stars are confined to a narrow range of 
$T_{eff}$ (e.g., Papers XIV, XVI, XVII, this study). Not only do RAC's vary 
with the spectral type (e.g. Stepien, 1989, 1993, 1994, Houdebine et al., 
2016), but also the distributions of rotation periods varies from one 
spectral sub-type to another (HM and this paper). 

Based on our previous papers, we have found that the most suitable initial 
selection parameter when we wish to identify a homogeneous sample of K or M 
dwarfs belonging to a specific sub-type is the (R-I) color: this color is 
sensitive to $T_{eff}$, but less so to metallicity (e.g. Leggett 1992, Ramirez 
\& Melendez 2005). Moreover, broad-band colors of high precision are widely 
available in the literature for many of the cool dwarfs which are of interest 
to us.

\subsection{Selection of a sample of late K dwarfs}

We selected a sample of 419 late K dwarfs on the basis of (R-I) measurements 
available in the literature. Our sample contains stars with (R-I)$_{C}$ 
(i.e. (R-I) color in the Cousins system) in the range [0.684;0.816] which 
also corresponds to (R-I)$_{K}$ ((R-I) in the Kron system) in the range 
[0.503;0.613] according to the transformation formulae of Leggett (1992) (see 
Leggett 1992 for more information on the Cousin's and Kron photometric 
systems). According to Kenyon and Hartmann (1995), this range of colors is 
centered on (R-I)$_{c}$=0.75, i.e. the spectral type dK7. However, when we 
compiled and derived effective temperatures with the PCA-based inversion 
method (see Paletou et al. 2015 and Sect.~3) for this sample of late K 
dwarfs, we found in average higher temperatures than what would be expected 
from the (R-I)$_{C}$-$T_{eff}$ tabulation of Kenyon and Hartmann (1995). We 
give the stellar parameters and spectral types of our late K dwarf sample in 
Table~1. One can see in this table that most stars are dK5 or dK6 stars, 
with a few dK4 and dK7 stars. In view of this relatively large range of 
spectral types, we decided to restrict our sample to the range dK5.7 to dK7.3 
in order to obtain a better defined RAC (Houdebine et al. 2016). This 
restricted sample is centered on 
the spectral sub-type dK5.9. We shall therefore refer to this sample as the 
dK6 stars. Our dK6  stellar sample contains stars that have similar (R-I) 
colours and the same effective temperatures to within $\pm$110 K. We refer to 
HM for a discussion of corresponding data for our sample of dM3 stars. Values 
of (R-I)$_{C}$ for our samples of stars were taken from the following papers: 
Veeder (1974), Eggen (1974), Rodgers \& Eggen (1974), Eggen (1976a, 1976b), 
Eggen (1978), Eggen (1979), Eggen (1980), Weis \& Upgren (1982), Upgren 
\& Lu (1986), Eggen (1987), Booth et al. (1988), Leggett \& Hawkins (1988), 
Bessel (1990), Weis (1991a 1991b), Dawson \& Forbes (1992), Leggett (1992), 
Weis (1993). 

These papers provided us with a starting list of a large number (419) of late 
K dwarfs.  Searching through databases at the European Southern Observatory 
(ESO) and Observatoire de Haute Provence (OHP), we identified spectra of 112 
different stars which are suitable for our purposes. The final list of 105 
stars in our sample late K dwarf sample is provided in Table~1 (note that in 
Table~1 we also added other stars with known $v\sin i$ and [M/H], see next 
sections).
 
The spectra which we use for determining $v\sin i$ came from two different 
\'echelle spectrographs;  HARPS (High Accuracy Radial velocity Planet Search, 
ESO) and SOPHIE (OHP). These two high-resolution \'echelle spectrographs are 
designed for planet search programs and are very stable in wavelength and 
resolution. The instrumental stability makes the spectra well suited for the 
measurement of rotation in M dwarfs (see HM). We included in our sample the 
SOPHIE observations obtained in the High Efficiency (HE) mode for the purpose 
of measuring the Ca\,{\sc ii} and H$_{\alpha}$ line equivalent widths 
(Houdebine et al. 2016). For further details of the 
spectrographs, see HM. For measuring $v\sin i$ with SOPHIE spectra, we used 
observations in the High Resolution (HR) mode only (R=75000 and an image 
scrambler).

For comparison, when we were studying dM3 stars (see HM), we started off with 
a list of 381 dM3 objects based on (R-I)$_C$ data in the literature. Searching 
through the same databases as above, we found suitable observations for 86 
different dM3 stars.

\subsection{Biases in our sample stars}

The stars in our sample includes all stars from all observing programs 
which have been carried out with HARPS and SOPHIE. These programs are mostly 
planet-search programs, although some programs are dedicated to magnetic 
activity. Because of requirements of achieving high S/N in the spectra, 
the selection of stars is brightness limited.

Our initial target list of 419 late K dwarfs (based on broad-band colors) is 
estimated to be complete down to apparent brightness $m_V$ $\approx$ 11. In 
our sample of 105 stars with sufficiently high S/N spectra for our present 
purposes, the list is estimated to be complete down to $m_V$ $\approx$ 10 for 
the slow rotators. Assuming an absolute magnitude $M_V$ $\approx$ 8.5 as the 
boundary between normal disk stars and subdwarfs, our sample of 105 late K 
dwarfs is estimated to be complete out to a distance of 20 pc for our slow 
rotators. For subdwarfs, we cannot claim that our target list is complete.

As far as the bias towards bright stars at spectral type dM3 is concerned, HM 
reported that this bias has little effects on the mean rotation period of 
low-activity M dwarfs. This is due to the fact that the rotation period does 
not change much with the stellar radius for disk stars (see HM). 
Analogously, the brightness-limited bias in our late K dwarf sample is 
expected to have only minor effects in the RAC. The bias may contribute 
somewhat to the density of sampling at different parts of the RAC, but this 
is not expected to cause significant discrepancies as regards the overall RAC 
(Houdebine et al. 2016).

In the planet-search programs, observers tend to avoid stars with high levels 
of magnetic activity: spot modulation in such stars can mimic the effects of 
planets and can also  add significant noise to the data. Therefore our spectral
 samples are likely to be biased in general towards low activity stars, 
i.e. stars which are far removed from ``saturated" magnetic effects. Despite 
this ``anti-activity" bias, we found that a small percentage of more active 
(i.e. dMe) stars did survive in our samples: 2\% of our late K dwarf sample 
are catalogued as dKe, and 10\% of our dM3 sample are catalogued as dM3e. We 
note that the latter figure is not far from the report of West et al (2008) 
that 10-20\% of dM3 stars are dM3e. However, we completed our own samples 
with $v\sin i$ measures and Ca\,{\sc ii} and H$_{\alpha}$ line equivalent 
width measures from the literature, including notably many new fast rotators. 
Therefore, the samples of stars studied in this paper do not contain {\it 
only} low activity stars (i.e. slow rotators).

\begin{figure*} 
\vspace{-1.5cm}
\begin{centering}
\hspace{-1.5cm}
\includegraphics[width=14cm,angle=-90]{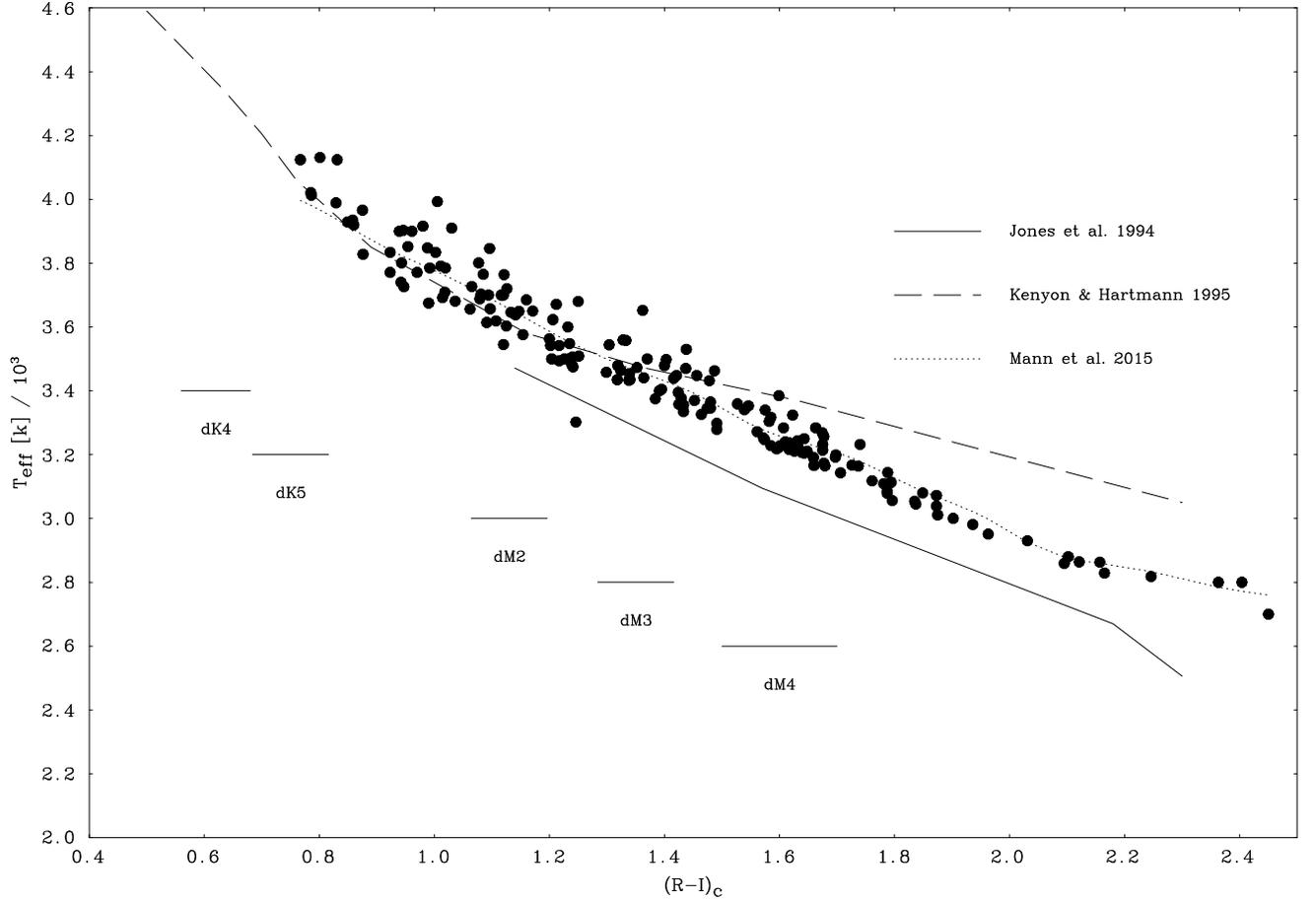}
\vspace{-0.5cm}
\end{centering}
\caption[]{Values of $T_{eff}$ as a function of the (R-I)$_C$ color for the 
data of Mann et al. (2015, filled circles) and a smoothing of this data (dotted
 line). Also shown for comparison: the values of Jones et al. (1994, solid 
line) used in some of our previous studies. A difference of about $\sim$170~K 
exists between the two correlations. Also shown: the calibration of Kenyon and 
Hartmann (1995). This later calibration agrees well with the more recent 
measures of Mann et al. up to (R-I)$_C$=1.3, but tends to overestimate 
$T_{eff}$ for later spectral types. We overplot the (R-I)$_C$ domains of our 
samples of stars from dK4 to dM4.}
\end{figure*}

\section{Stellar parameters}

Here, we discuss the methods we use to derive $T_{eff}$ and radius for each of 
our samples of stars. The radius is necessary for the determination of the 
projected rotation periods, P/sini. 

\subsection{PCA-based inversion of stellar fundamental parameters for late K 
dwarfs}

In order to derive estimates of effective temperatures ($T_{eff}$) for our 
late K dwarfs In Table~1 we used the calibration of Kenyon \& Hartmann (1995: 
hereafter KH). We derived the effective temperature as a function of 
(R-I)$_{C}$ according to this tabulation. We searched the literature for other 
$T_{eff}$ measurements for our late-K targets (we used the tutorial developed 
by Paletou \& Zolotukhin 2014\footnote{http://www.astropy.org/}). We found a 
relatively good agreement between these temperatures and effective 
temperatures previously published: The mean difference in only of the order 
of 75~K.

As an independent method of determining values of $T_{eff}$ for our sample 
of late K dwarfs, we used the principal component analysis-based (PCA) method 
to derive T${eff}$ and also [Fe/H], $log(g)$ and $v\sin i$ (Paletou et al. 
2015b). HARPS and SOPHIE spectra  were inverted using the PCA method introduced
 for stellar fundamental parameters by Paletou et al. (2015a). In the latter 
paper, stellar spectra of FGK stars were inverted using a so-called ``learning 
database" made from the Elodie stellar spectra library (see Prugniel et al. 
2007): the ``learning" occurs by using \emph{observed} spectra for which 
fundamental parameters had already been evaluated. We note that the spectra 
considered in Paletou et al. (2015a) had typical spectral resolutions 
${\cal R}$ of 50\,000 and 65\,000: these are significantly lower than the 
resolution of HARPS and SOPHIE data which will be analyzed in the present 
paper.

For the present study, we used the PCA method, but with a difference: the 
``learning" was achieved by referring to \emph{synthetic} spectra. A grid of 
6336 spectra was computed using {\sc Synspec}-48 synthetic spectra code
(Hubeny \& Lanz 1992) and Kurucz {\sc Atlas}-12 model atmospheres (Kurucz 
2005). The linelist was built from Kurucz (1992)
{\tt gfhyperall.dat}\footnote{http://kurucz.harvard.edu}. In order to deal 
with stars close to dK6, we adopted a grid of parameters such that $T_{eff}$ 
is in the range 3500--4600 K (with steps of 100 K), log$g$ is in the range 
4--5 cgs (with steps of 0.2 dex), metallicity [Fe/H] is in the range from -2 
to +0.5 (with steps of 0.25 dex) and, finally, $v{\rm sin}i$ is in the range 
from 0 to 14 $km\,s^{-1}$ (with steps of 2 $ {\rm km\,s}^{-1}$). For all 
models the microturbulent velocity was fixed at $\xi_t = 1\ {\rm km\,s}^{-1}$ 
and [$\alpha$/Fe] was set to 0. Also, for producing the reference spectra, we 
adopted the spectral resolution of the HARPS or SOPHIE spectrograph i.e., 
${\cal R} = 115\,000$ or ${\cal R} = 75\,000$. We finally limited the PCA 
study to a spectral band almost 100~\AA\ wide centred on the Na\,{\sc i} 
D-doublet, ranging from 585.3 to 593.2 nm.

Even though the choice of {\sc Atlas}-12 for such cool stars, as well as the 
choice of the synthetic spectral band we considered may be open to debate, we 
believe that parameters inverted from HARPS and SOPHIE data using our PCA 
approach are realistic enough for an initial analysis (see Paletou et al. 
2015b). 

Table~2 provides the results we have obtained from our PCA approach to the 
sample of 112 late K dwarfs as regards $T_{eff}$, log$g$, [Fe/H], and $v\ sin 
i$. Typical uncertainties for these four parameters are respectively of order 
100~K, 0.2 dex, 0.1 dex, and 1.5 $ {km\,s}^{-1}$. It is important to note that 
the $v\ sin i$ values derived from PCA are only preliminary estimates in order 
to obtain synthetic spectra to be used for ``learning". Later on (Section 4 
below), we shall derive more precise $v\ sin i $ values based on the 
cross-correlation analysis: then we shall compare (Section~5) the precise 
values with the estimates derived from PCA (see Figure~12 below). Note that 
the spectra of GJ~1248 and Gl~747.1 were not inversed with the PCA method 
because they were too noisy.

In our sample of 112 stars, the PCA approach indicated that seven stars have 
significantly lower $T_{eff}$ values than those expected for a late K dwarf: 
the seven stars are GJ~1248,  Gl~17.1, Gl~369, Gl~401, Gl~747.1, Gl~842, and 
Gl~855 (see Table~2). It appears that our sample of late K dwarfs is polluted 
by a few cooler stars with spectral types which are actually later than dK7. 
Confirmation of the unusual coolness of 7 stars in our list of 112 is provided 
by our PCA determinations of $T_{eff}$ in Table~2. Clearly, some of the 
(R-I)$_{C}$ values in our sample are discrepant: In view of this, we have 
excluded the 7 discrepant stars from our initial sample of 112 objects 
(Table~1).

\begin{figure*} 
\vspace{-1.5cm}
\begin{centering}
\hspace{-4.5cm}
\includegraphics[width=10cm,angle=-90]{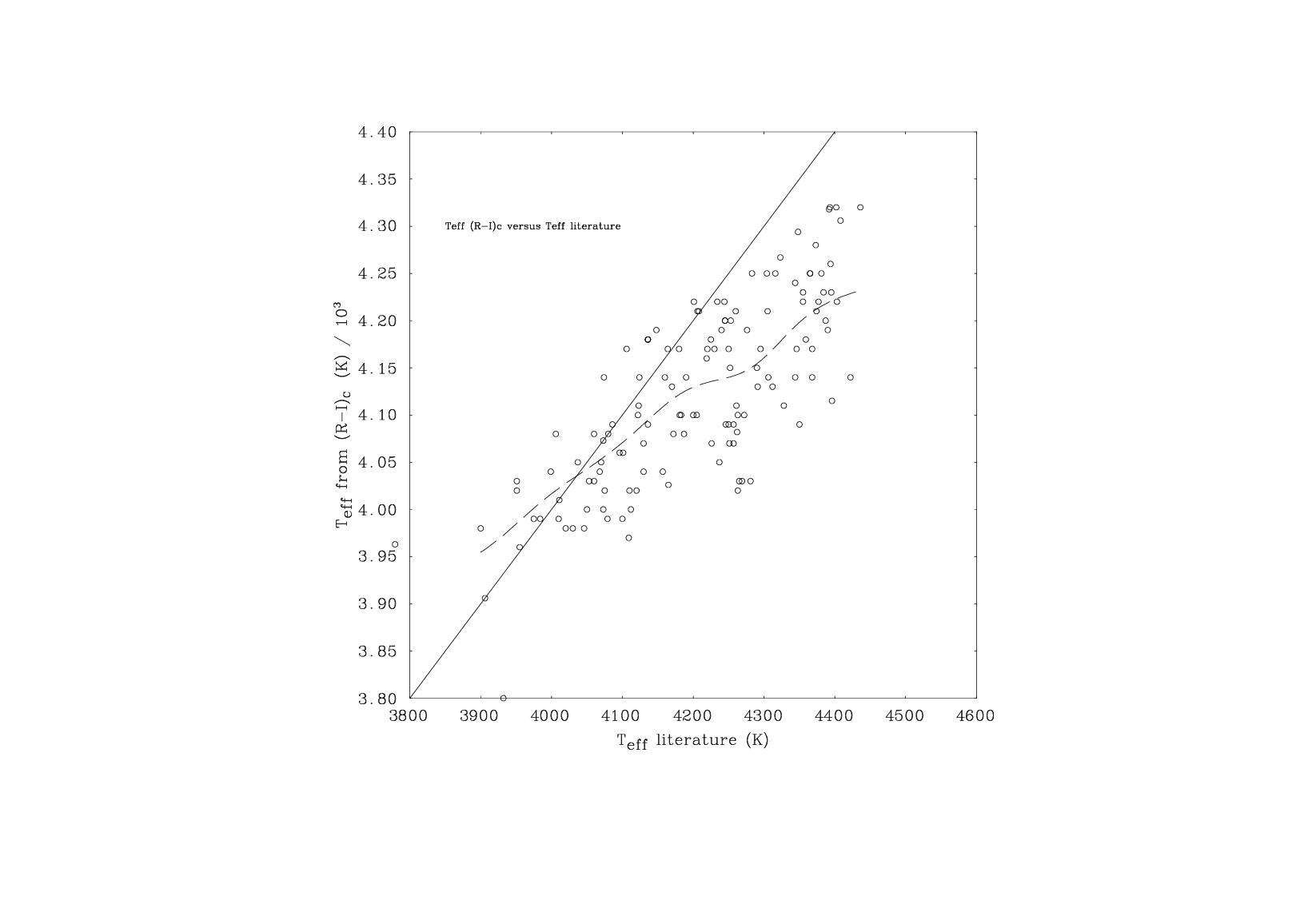}
\vspace{-0.5cm}
\hspace{-4.5cm}
\includegraphics[width=10cm,angle=-90]{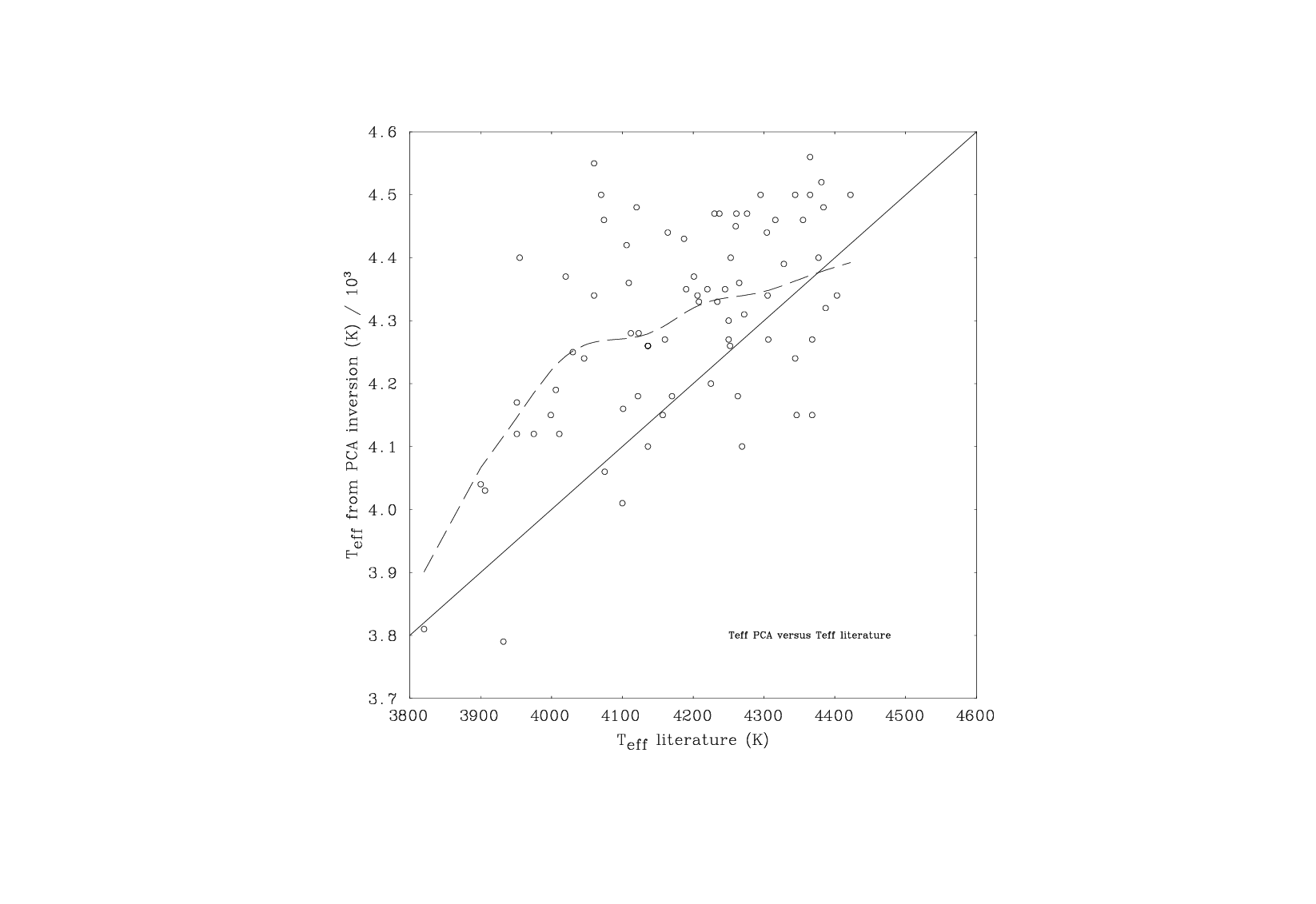}
\end{centering}
\caption[]{Values of $T_{eff}$ derived from the (R-I)$_C$ colour as a function 
of the $T_{eff}$ values compiled from the literature (left panel). The solid 
line is the one-to-one correlation. The dashed line shows the smoothed data. 
Right Panel: Values of $T_{eff}$ derived from the PCA inversion method as a 
function of the $T_{eff}$ values compiled from the literature. Same description
as the left panel. }
\end{figure*}

Now let us compare the results for $T_{eff}$ which we have derived by means of 
the two different methods: KH and PCA. For our sample, we obtained a mean 
effective temperature of 4130~$\pm 100$K when we use the (R-I)$_C$ colour 
calibration method (KH). However, for the same sample of stars we obtained a 
mean effective temperature of 4270~$\pm 100$K when we use the PCA approach. 
There is a systematic difference of 140~K between PCA and colour calibration. 
Comparatively, the (R-I)$_C$ colour calibration method (KH) is in better 
agreement with the values given in the literature (see references therein), 
with a systematic difference of only 75~K. We show in Fig.~2 the values of 
$T_{eff}$ derived from the (R-I)$_C$ colour as a function of the $T_{eff}$ 
values compiled from the literature (left-hand panel). The solid line is 
the one-to-one correlation. The dashed line shows the smoothed data. There 
is a significant scatter among the data: typically $\pm 70$~K. Our measures 
inferred from (R-I)$_C$ agree rather well with other published values, 
typically within 50~K, up to a temperature of 4200~K. For larger values our 
temperature inferred from (R-I)$_C$ tend to be lower than published values, 
up to a value of 100~K. However, this latter difference is still reasonable. 
We can say that globally the temperatures inferred from (R-I)$_C$ are good 
evaluations. Houdebine et al. (2016, in preparation) show that (R-I)$_C$ is a 
good effective temperature diagnostic (based on the calibration of Boyajian et 
al. 2012) from spectral types dK3 to dM7 with typical differences with 
a large compilation of effective temperatures from the literature of only 
$\pm 40$~K.

We show in Fig.~2 (right-hand Panel) the values of $T_{eff}$ derived from the 
PCA inversion method as a function of the $T_{eff}$ values compiled from the 
literature. One can note that the PCA method tends to globally over-estimate 
$T_{eff}$ at low and large values. The maximum difference in average is of 
the order of 200~K. 

The differences we find between the different approaches are to be compared 
with systematic difference of typically 200-300~K between the results obtained 
by various investigators for K dwarfs (e.g. Ramirez \& Melendez 2005; Ammons 
et al. 2006; Morales et al. 2008; Jenkins et al. 2009; Soubiran et al. 2010). 
We found during the course of our compilation of $T_{eff}$ from the literature 
for our objects that differences between authors are often different by more 
than 100-200~K. In some cases, the differences may attain more than 500~K. 
These systematic differences seem relatively large compared to the precisions 
which are claimed by the various authors (typically $\pm$100 K). In view of 
these systematic differences already in the literature, we believe that the 
$T_{eff}$ differences which we encounter between our two methods are not 
excessive: in fact, given the scatter in the data, we consider our results 
reasonable and consistent with values in the literature. In what follows, we 
have taken the mean of the three different values (KH, PCA and mean of 
published values) in order to obtain reliable estimates of our effective 
temperatures for our sample of late K dwarfs (Table~1).

According to our PCA approach, the mean metallicity of our sample is -0.316 
dex (excluding dKe stars and subdwarfs). Therefore, on average, our late K 
dwarfs are metal poor. However, if we compare this value to the average of the 
metallicities for dM2 stars (-0.154 dex, Paper XIV) we find that the present 
metallicities for our sample may be underestimated. As a matter of fact, for 
dM2 stars, we obtained a good RAC after correction from the metallicity 
effects in Paper XIV, which somewhat confirms a posteriori the accuracy of 
those metallicities. On the other hand, when we attempt to correct the dK6 
star RAC for metallicity effects (Houdebine et al. 2016), we obtain a larger 
scatter in the RAC. This may arise from the larger uncertainty we have on the 
metallicity for dK6 stars (0.2 dex).

\subsection{Effective temperatures for our samples of M dwarfs and the Stellar 
radii}

In order to derive radii for the stars in our dK6 and other samples, we used 
the classical formula;

\begin{equation}
M_{v}+BC = 42.36 - 5\times log\frac{R_{*}}{R_{\odot}} - 10\times log T_{eff},
\end{equation}

\noindent
where symbols take their usual meaning. We used the BC tabulation as a 
function of (R-I)$_C$ from KH: these authors compiled data on BC as a function 
of various colours from the literature. We estimate that their tabulation 
yields an uncertainty of about 10\% in BC at spectral type K6 (see KH). Using 
Eq. (1), we obtained the radii for our late K dwarfs as listed in Table 1. In 
this table, we give the mean of the effective temperatures derived from the 
(R-I)$_C$ colour from KH, the values of the PCA method and values previously 
published. The sources of the published effective temperature are: Wright et 
al. (2003), Valenti \& Fischer (2005), Ammons et al. (2006), Masana et al. 
(2006), Sanchez-Blazquez et al. (2006), Sousa et al. (2006), Cenarro et al. 
(2007), Morales et al. (2008), Jenkins et al. (2009), Lafrasse et al. (2010), 
Soubiran et al. (2010), Casagrande et al. (2011), Prugniel et al. (2011), 
McDonald et al. (2012), Chen et al. (2014), Gaidos et al. (2014). We estimate 
the errors as the $\sigma$ of these measures. The errors on the radii in 
Table~1 were computed taking into account the combination of the error on the 
absolute magnitude, an assumed error of 0.02 magnitudes on the (R-I)$_C$ 
colour, and a typical 100 K error in $T_{eff}$. The mean $T_{eff}$ and radius 
for our dK6 sample are 4226~K and 0.6469 R$_{\odot}$ respectively. The value 
of the radius for this $T_{eff}$ given by Boyajian et al. (2012) is 
0.6457~R$_{\odot}$. Therefore the two values agree within 0.19\%. We generally 
considered that in our dK6 sample, stars with [M/H]$<$-0.5 ($M_{V}\approx 8.6$)
 are subdwarfs. 

\begin{figure*} 
\vspace{-0.5cm}
\hspace{-2.5cm}
\includegraphics[width=14cm,angle=-90]{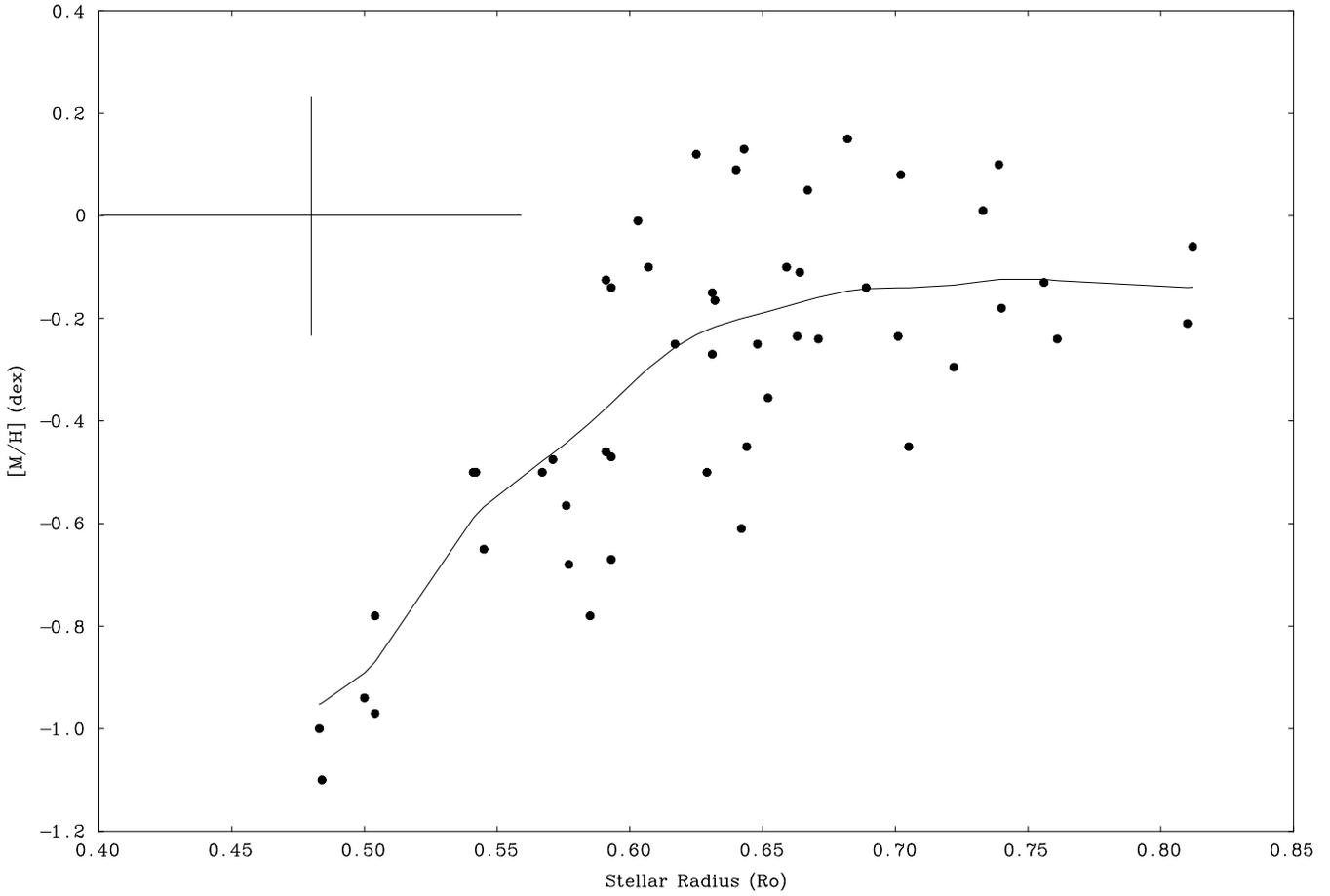}
\vspace{-0.5cm}
\caption[]{Correlation between the metallicity [M/H] and the stellar 
radius for stars with spectral sub-type dK6. The two parameters do correlate 
for dK6 stars but the scatter is large. This large scatter is essentially due 
to the large errors on the stellar radii and [M/H] as indicated by the error 
box.}
\end{figure*}

For our samples of dM2 and dM4 stars, we used in previous studies the 
calibration of Jones et al. (1994). However, a more recent study which 
should be more precise (Mann et al. 2015) indicates that the values from 
Jones et al. (1994) may be underestimated by about 170~K. We show the 
calibrations of $T_{eff}$ as a function of (R-I)$_C$ of Jones et al. (1994), 
KH and Mann et al. (2015, filled circles) in Fig.~1. We also show a smoothing 
of the data of Mann et al. (2015, dotted line). The calibration of Kenyon and 
Hartmann (1995) agrees well with the more recent measures of Mann et al. 
(2015) up to (R-I)$_c$=1.3, but tends to overestimate $T_{eff}$ for later 
spectral types. Therefore, the stellar radii have been overestimated in our 
previous studies of dM2 and dM4 stars by the order of 10-15\%. 

In the coming sub-sections, our measurements of $v\sin i$ have been compared 
with, and supplemented by, those from other authors. For comparison between 
the different spectral sub-types 
used in this study, we overplot the (R-I)$_C$ domains of our samples of stars 
from dK4 to dM4. However, we recall that there are systematic differences of 
the order of 200-300~K in the $T_{eff}$-radius plane between the 
interferometric calibrations of Boyajian et al. (2012) and the 
Double-Eclipsing Binaries calibrations (see Boyajian et al. 2012 for a 
comparison) for K and M dwarfs. Therefore, the calibration of Mann et al. 
(2015) may not be definitive and may still be subject to improvements.

We give in Tables 3, 4 and 5 the stellar parameters for our samples of dM2, 
dM3 and dM4 stars respectively. The mean $T_{eff}$ and radii for our samples 
of dK4, dK6, dM2, dM3 and dM4 stars are given in Table~6. We find values of 
4600~K, 4226~K, 3658~K, 3462~K and 3317~K for the mean $T_{eff}$ of our 
samples of dK4, dK6, dM2, dM3 and dM4 stars respectively. The mean radii 
are 0.6469~R$_{\odot}$, 0.5082~R$_{\odot}$, 0.4090~R$_{\odot}$ and 
0.3008~R$_{\odot}$ for the dK6, dM2, dM3 and dM4 samples respectively. 
These radii agree with those of Boyajian et al. (2012) within 0.19\%, 6.3\%, 
4.3\% and 8.2\% respectively, which is acceptable for the studies of the 
$P\sin i$ distributions and the RAC's.

For our sample of dK5 stars (Houdebine 2011a and 2012b, 
Papers XVI and XVIII), we found that the mean temperature is 4600~K (Table~6). 
Effective temperatures for these stars were computed from values published in 
the literature (see Paper XVI). The stars were initially selected with 
(R-I)$_C$=[0.56:0.68] which corresponds to the spectral sub-type $\sim$K5 
(KH). However, as in the case of dK6 stars, we found that the published 
effective temperatures were somewhat higher and rather correspond in average 
to the spectral sub-type dK4 (KH). Therefore, we shall refer to the dK4 stars 
for this sample of stars thereafter. We do not repeat the stellar parameters 
of our dK4 stellar sample here as the effective temperatures have not changed 
and the data is available in Paper XVI.

\subsection{Stellar metallicities}

\begin{figure*} 
\vspace{-0.5cm}
\hspace{-2.5cm}
\includegraphics[width=14cm,angle=-90]{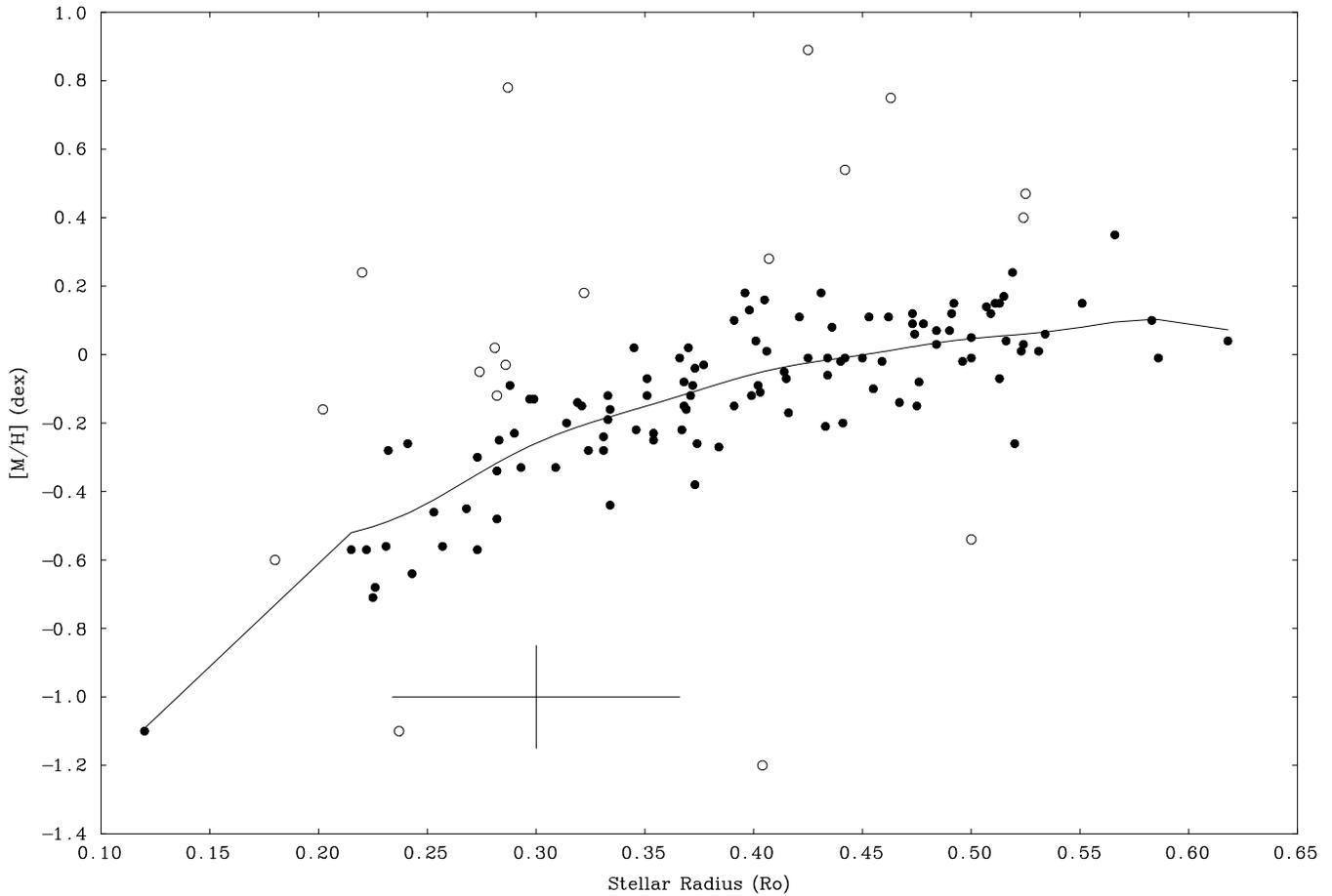}
\vspace{-0.5cm}
\caption[]{Correlation between [M/H] and stellar radius for stars with 
spectral sub-type dM3. The correlation is relatively tight as in dM2 stars 
(Paper~VII) in spite of the relatively large error on the radius (error box). 
We can see that, as in dM2 and dK6 stars, the metallicity decreases with 
decreasing radius for a given (R-I)$_{C}$. We show some obvious outliers as 
open circles and the continuous line represents the data smoothed by a 
Gaussian.}
\end{figure*}

We saw in Paper~XV that correction from metallicity effects
 in dM2 stars is essential in order to obtain a good correlation 
between the Ca\,{\sc ii} EW and $P/\sin i$. This is due to the fact that 
the Ca\,{\sc ii} line formation depends sensitively on the Ca abundance 
(Houdebine \& Panagi, in preparation). The same applies to dK6 stars 
and to the Ca\,{\sc ii} surface fluxes. Therefore, we compiled [M/H] measures 
from the literature for our dK6 stellar sample in order to obtain a 
metallicity-radius empirical correlation as in Houdebine (2008, Paper VII), 
and correct our Ca\,{\sc ii} surface fluxes from metallicity effects, 
assuming a proportionality between [M/H] and Ca\,{\sc ii} surface fluxes. 
We found several measures of [M/H] from the following authors: Marsakov \& 
Shevelev (1988), Alonso et al. (1996), Valenti \& Fischer (2005), Ammons et 
al. (2006), Sanchez-Blazquez et al. (2006), Sousa et al. (2006), Cenarro et 
al. (2007), Mishenina et al. (2008), Soubiran et al. (2010), Casagrande et al. 
(2011), Milone et al. (2011), Prugniel et al. (2011), Wu et al. (2011), 
Anderson \& Francis (2012), Gaidos et al. (2014). We averaged the values from 
the literature with those from our PCA-based inversion method.

We plot in Fig.~3 these average values of [M/H] as a function of radius for 
our dK6 stars. One can see in this diagram that the scatter is large. We 
show the error box in Fig.~3, which gives in [M/H] the mean of the 
differences between values from the literature and our values from the 
PCA-based inversion method, and in radius, the mean of the errors on the 
radii for our stars. Both errors are large and explain the poor 
correlation between these two parameters for dK6 stars as well as the large 
scatter at about the running mean (solid line). As a consequence, the 
corrections from metallicity turn out to be rather poor: in fact, we do not 
obtain an improved RAC after corrections for [M/H] compared to the RAC without 
 such corrections (see Houdebine et al. 2016).

For our dM2 stellar sample, we used the empirical correlation found by 
Houdebine (2008, Paper VII) to determine the metallicity for each stars as 
a function of its radius. These values were computed in Paper~XV and are 
reported in Table~3. 

We show the correlation between [M/H] and stellar radius for stars with 
spectral sub-type dM3 in Fig.~4. The correlation is relatively tight as in 
dM2 stars (Paper~VII) in spite of the relatively large error on the radius 
(see the error box in Fig.~4). The scatter in this correlation is about three 
times less than in the same correlation for dK6 stars (Fig.~3). We can see 
that, as in dM2 and dK6 stars, the metallicity decreases with decreasing 
radius for a given (R-I)$_{C}$. We show some obvious outliers as open circles 
and the continuous line represents the data smoothed by a Gaussian. We 
derived the metallicities for dM3 stars according to this mean correlation. 
We report the values of [M/H] in Table~6.

We decided also to compile all the metallicities published for our initial 
selection list of 395 dM4 stars, and try to obtain a metallicity-radius 
correlation for these stars. We found metallicities from the literature for 
179 dM4 stars. We 
give these metallicities in Table~5. Our measures of [M/H] for our list of 
dM4 stars were collected from the following authors: Marsakov \& Shevelev 
(1988), Ivanov et al. (2004), Ammons et al. (2006), Sanchez-Bazquez et al. 
(2006), Cenarro et al. (2007), Holmberg et al. (2009), Jenkins et al. (2009), 
Soubiran et al. (2010), Casagrande et al. (2011), Milone et al. (2011), 
Prugniel et al. (2011), Wu et al. (2011), Anderson \& Francis (2012), Koleva 
\& Vazdekis (2012), Rojas-Ayala et al. (2012), Kordopatis et al. (2013), Mann 
et al. (2013), Pace (2013), Gaidos et al. (2014), Gomez da Silva et al. 
(2014), Newton et al. (2014), Mann et al. (2015). We show the empirical 
relationship between [M/H] and stellar radius for stars with spectral 
sub-type dM4 in Fig.~5. 

\begin{figure*} 
\vspace{-0.5cm}
\hspace{-2.5cm}
\includegraphics[width=14cm,angle=-90]{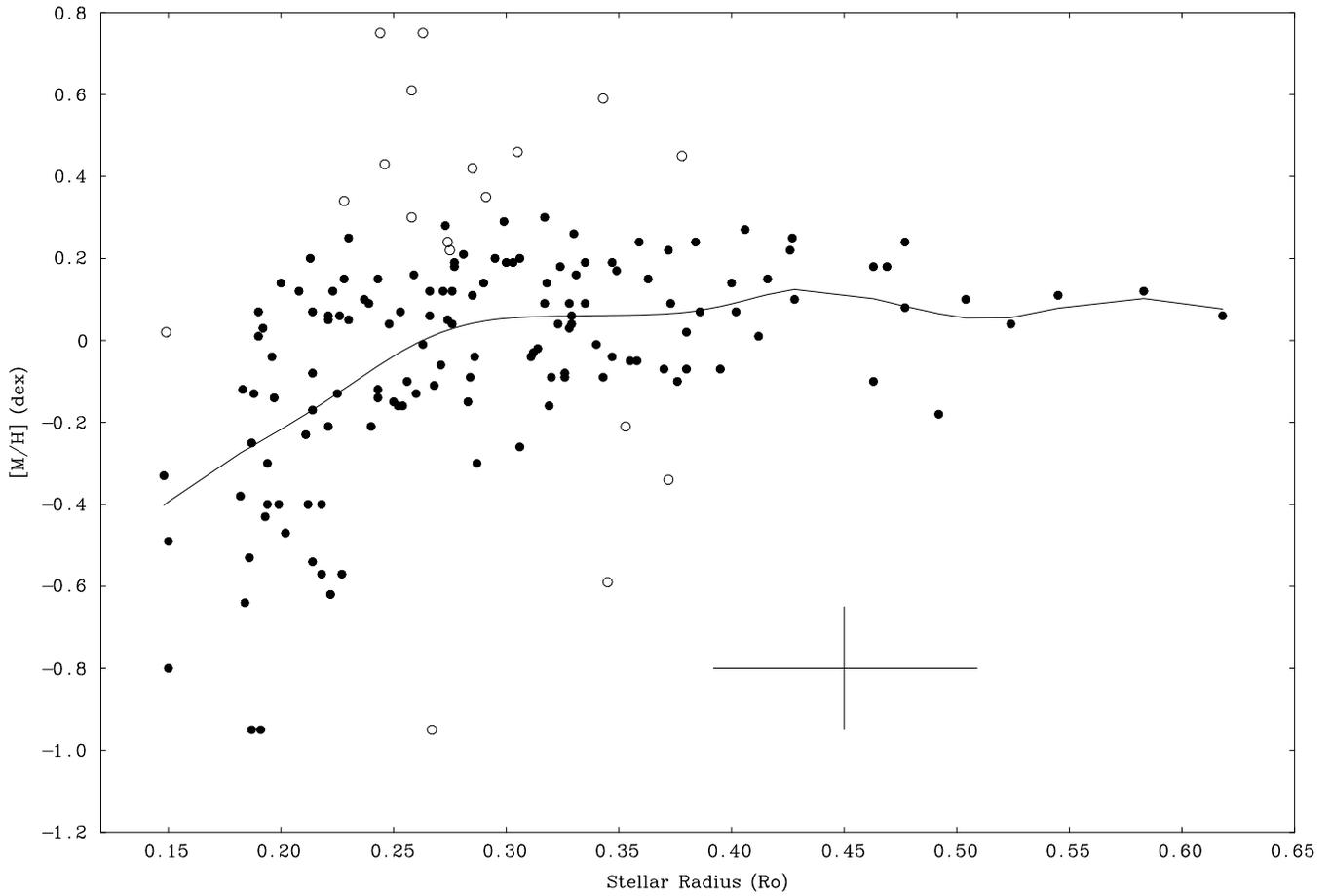}
\vspace{-0.5cm}
\caption[]{Correlation between [M/H] and stellar radius $R_{*}$ for stars with 
spectral sub-type dM4. The open circles are outliers that were not included 
in the smoothing of the data (solid line).}
\end{figure*}

The correlation is not as tight as in dM2 stars or dM3 stars, 
particularly for small radii where the scatter is large (about $\pm 0.5$ dex). 
We have no simple explanation for this relatively large scatter at this stage. 
We also note that the mean of the data (continuous line) is rather flat and 
stands at [M/H] of about 0 dex over a large range of radii. This was not the 
case for dK6, dM2 and dM3 stars for which [M/H] starts to diminish more 
rapidly at small radii. In Fig.~5, we show some obvious outliers as open 
circles and the continuous line represents the data smoothed by a Gaussian. 
Also, for dM4 stars, the minimum values of [M/H] do not fall below -0.5 dex 
for the smoothed curve even for the smallest radii which questions the 
presence of subdwarfs in our sample, in spite of the fact that individual 
measures of [M/H] include about ten stars below -0.5 dex (Table~7). We give 
the [M/H] values from the smoothed data in Table~7.

\section{Rotation of the dK6 stars}

\begin{figure*} 
\vspace{-1.5cm}
\begin{centering}
\hspace{-6.5cm}
\includegraphics[width=20cm,angle=-90]{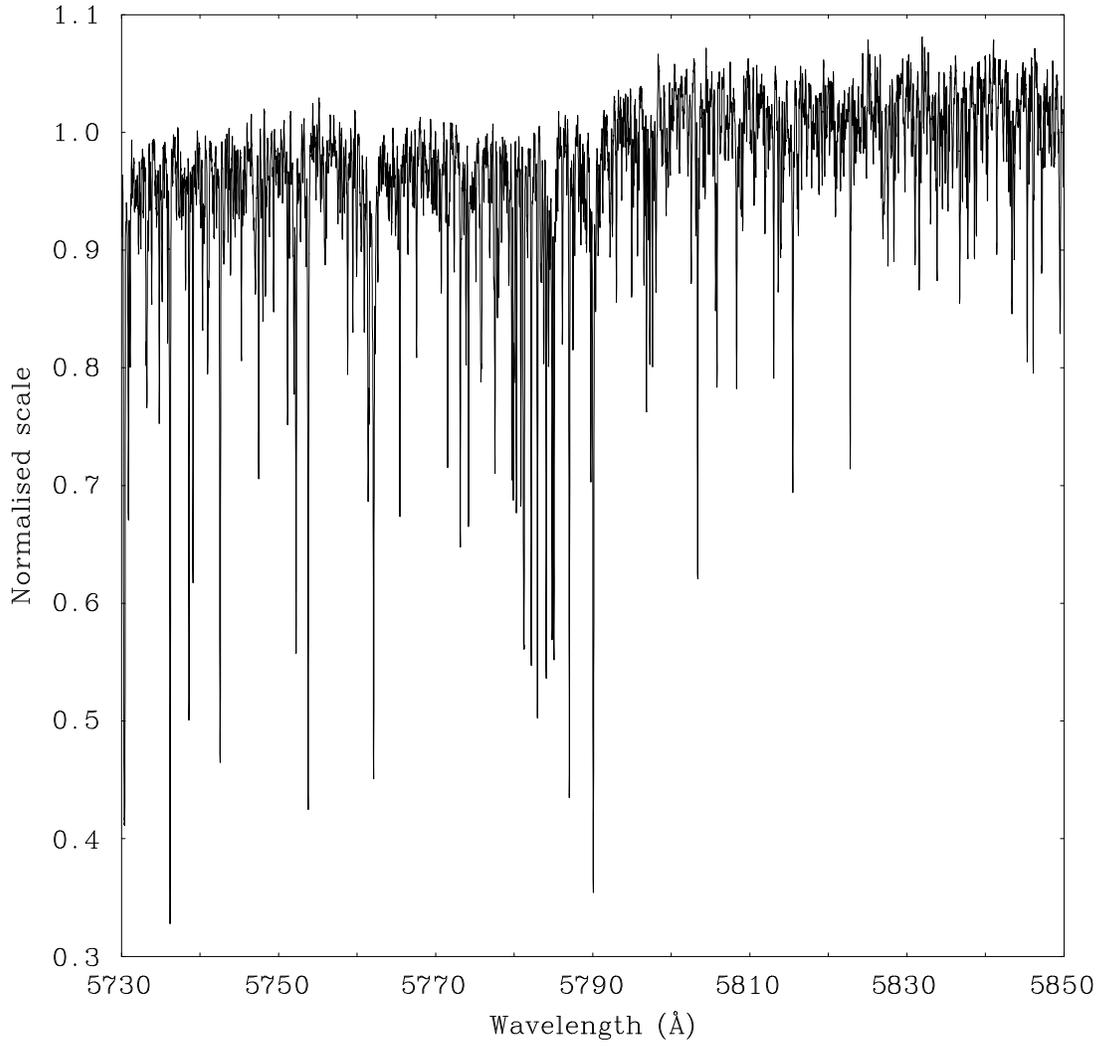}
\vspace{-0.5cm}
\end{centering}
\caption[]{A portion of the wavelength range used for our cross-correlation 
analysis of the late-K dwarf Gl 798: 5730\AA-5850\AA. This range yields very 
clean cross-correlation profiles for late-K dwarfs. For Gl 798, the S/N ratio 
is $\approx$ 380: the variations in the spectrum near the continuum level are 
actually numerous blended weak absorption lines, each of which contributes to 
the cross-correlation function for this star.}
\end{figure*}

\begin{figure*} 
\vspace{-0.5cm}
\begin{centering}
\hspace{-6.5cm}
\includegraphics[width=20cm,angle=-90]{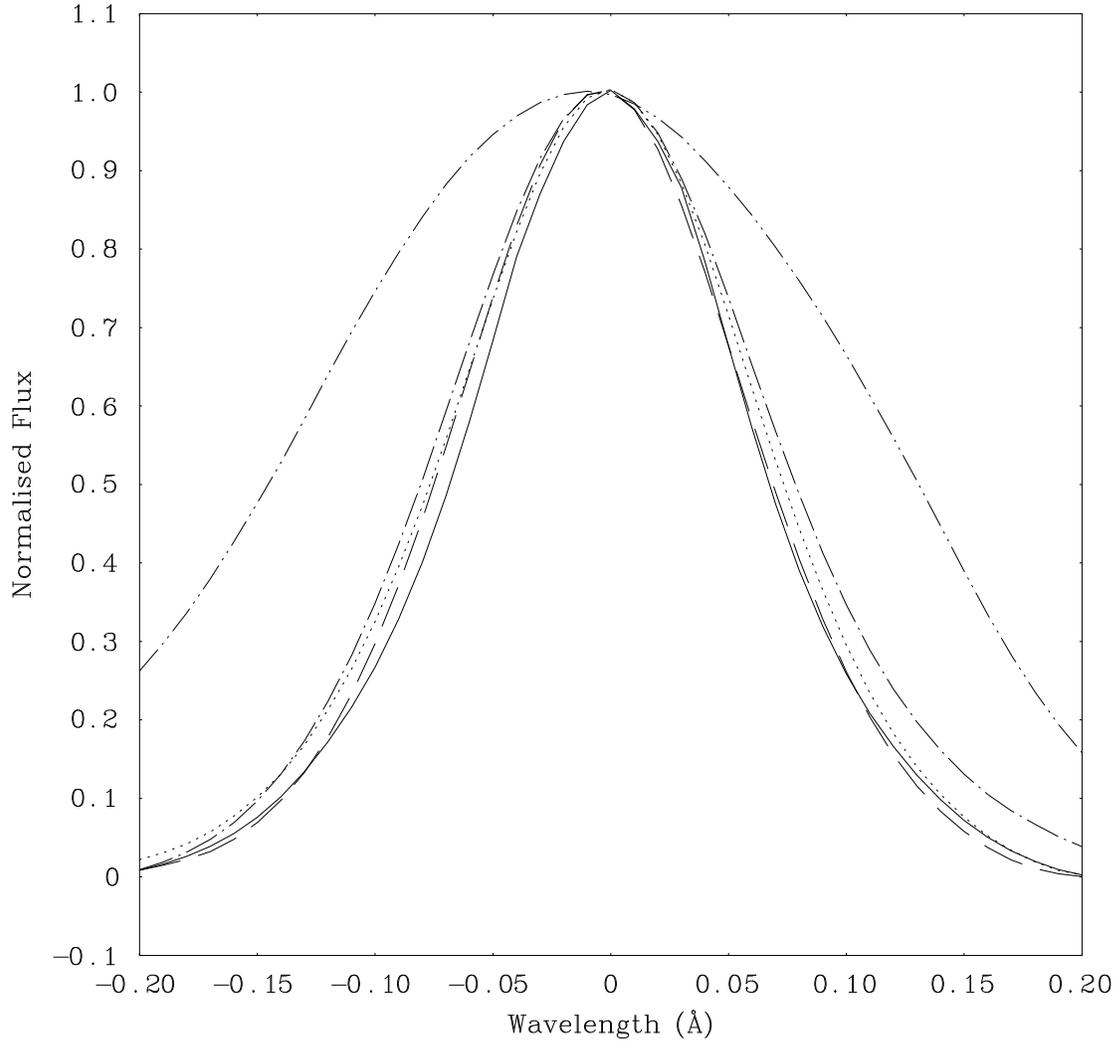}
\end{centering}
\vspace{-0.5cm}
\caption[]{Examples of cross-correlation functions obtained from our analysis 
of HARPS spectra of 5 late-K dwarfs in our sample:  Gl 57 (0 $km\ s^{-1}$), 
GJ~9299 (1.78 $km\ s^{-1}$), GJ 9714 (2.51 $km\ s^{-1}$), Gl 425B (3.35 $km\ 
s^{-1}$) and Gl 517 (9.77 $km\ s^{-1}$) (from narrowest to broadest). The FWHM 
of the peak for any particular star (in units of \AA ) is used to determine 
the value of $v\sin i$ for that star.}
\end{figure*}

In view of the inhomogeneity of our sample of 105 late K dwarfs, that include 
severall K4, K5 and K6 dwarfs and a few K7 dwarfs, we selected a sub-sample 
of stars from K5.7 to K7.3, i.e. an effective temperature of 4226$\pm 110$~K.
For this sub-sample of K6 dwarfs, it is possible to inter-compare the spectra 
and identify rotational broadening effects in the absorption lines.  
The profiles of the photospheric lines depend on three main four parameters; 
rotational broadening, micro-turbulence, macro-turbulence and Zeeman 
broadening. Micro and macro turbulence (typically of order 1~$km\ s^{-1}$) are 
not expected to vary much among our sub-sample of dK6 stars. Metallicity plays 
a role in the line strengths but not on macro- and/or micro-turbulence. 
The molecular lines we use for our cross-correlations have Land\'e factors 
that are generally lower than those of atomic spectral lines (e.g. Afram \& 
Berdyugina 2015). Only two well defined molecular bands have larger Land\'e 
factors (e.g. Reiners \& Basri 2006): TiO at 7050\AA, and FeH 
at 9900\AA. For high activity stars, the Zeeman broadening effect could be 
significant even for small Land\'e factors. But considering the large $v\sin i$
for these stars ($6-18\ km\ s^{-1}$), the Zeeman broadening effect should be 
completely negligible.

In order to extract rotational information, we carefully selected the spectral 
domain that we used for the 
cross-correlations: in Paper XIV, in our study of dM2 stars, we chose a 
125\AA\ wide spectral domain between 5460\AA\ and 5585\AA. Here, in order to 
obtain cleaner cross-correlation functions for dK6 stars, trial and error has 
led us to prefer a 120\AA\ wide spectral domain between 5730\AA\ and 5850\AA. 
In this latter domain there are fewer weak and unsaturated narrow spectral 
lines that yield clean cross-correlation functions as far as the 
cross-correlation background is concerned (see Sect. 5). Unfortunately, in the 
present study, we cannot exploit spectral lines in molecular bands at the red 
end of the spectrum (such as we could do for dM3 and dM4 stars): in the warmer 
dK6 stars, the bands are not strong enough to be useful. This is unfortunate, 
since molecular bands give the bests results for the cross-correlations (see 
Paper XVII, HM). As we shall see below, the noise in the background is central 
to reaching the highest precision in the $v\sin i$ measurements. We show in 
Fig.~6 the spectral domain we have chosen for our rotational analysis of dK6 
stars, using as an example the star Gl 798: this star, with $m_V $ = 8.82, is 
neither the brightest nor the faintest star in our sample, which ranges from 
$m_V$ = 7.87 to 11.80. In Fig.~6, the level of the continuum has been 
normalised to 1. 

For each star in our sample, we cross-correlated the spectra with themselves 
in order to identify at first our ``zero-rotation" template stars. For each 
spectral sub-type, we chose our templates empirically as the star with the 
lowest FWHM of the cross-correlation peak among the sample of stars at that 
sub-type. We found that the best template for our dK6 stars in the HARPS 
database is Gl 57 with a FWHM = 0.1236\AA\ which corresponds to a 
Gaussian with a width of 6.40 $km\ s^{-1}$. This is slightly broader than the 
0.1006\AA\ FWHM (5.40 $km\ s^{-1}$) at 5585\AA\ which we measured in HARPS 
spectra of dM2 stars, and slightly narrower than the 0.1455\AA\ FWHM 
(7.54~$km\ s^{-1}$) at 5790\AA\ which we measured for the dK4 stars. This is 
consistent with a decrease in the widths of photospheric lines as we go to 
later spectral type among K and M dwarfs. Gl 57 lies at the lower end of our 
selection range. Our mean effective temperature for this star is 4040~K which 
implies a spectral type of K7.1. Our PCA based inversion for this star gives 
a temperature of 4030 K which yields also a spectral type K7.1. We trust more 
the effective temperatures than the (R-I)$_{C}$ colour because for the late K 
dwarfs, $T_{eff}$ may vary a lot for a given (R-I)$_{C}$, because the 
gradient of $T_{eff}$ versus (R-I)$_{C}$ increases a lot in late K dwarfs 
(see Fig. 1). Therefore, Gl 57 appears to agree with our selection of dK6 
stars. We emphasize, that during the course of this study we also investigated 
the cross-correlation with Gl~847A, a low $v\sin i$ star with a somewhat 
higher effective temperature ((R-I)$_C$=0.761 and $T_{eff}$=4110~K) and 
earlier spectral type (K6.7). But finally we found little differences in the 
$v\sin i$ measures between these two templates. We therefore conclude that 
Gl~57 is a reliable template although it has a bit lower effective 
temperature than the rest of the sample.

In SOPHIE spectra, we found that 
Gl~728, has the narrowest cross-correlation function: we used that as our 
template spectrum in analyzing SOPHIE spectra. Note that the cross-correlation 
of Gl~116 with Gl~728 yields a FWHM of 0.1583\AA\ that is a bit narrower than 
the self cross-correlation function of Gl~728 with itself (0.1611\AA ). The 
cross-correlation of Gl~116 with itself yields a FWHM of 0.1475\AA\ that is 
significantly smaller than those of most other dK6 stars. We believe this star 
has a later spectral type than dK6.


\begin{figure*} 
\vspace{-0.5cm}
\hspace{-2.5cm}
\includegraphics[width=14cm,angle=-90]{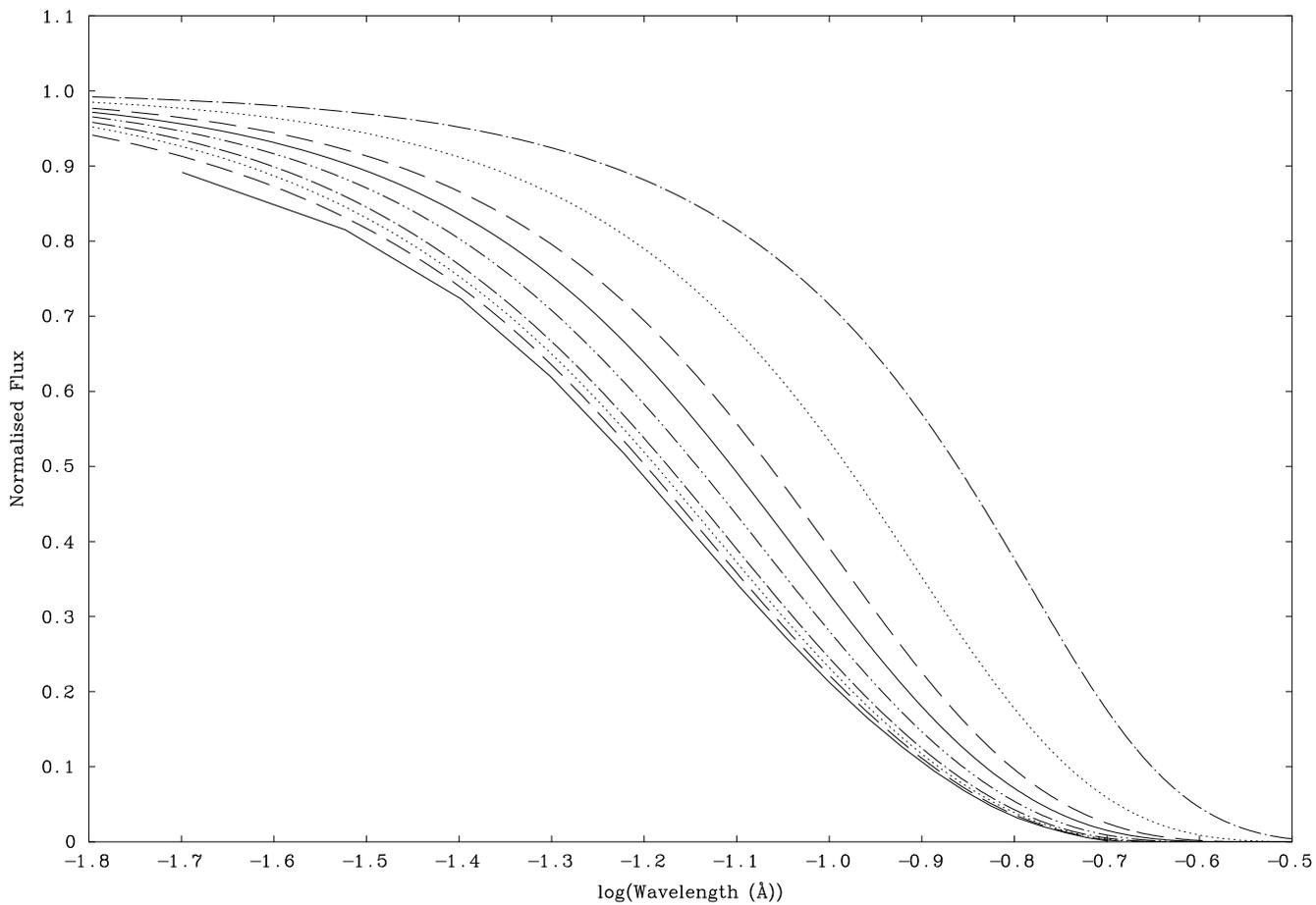}
\vspace{-0.5cm}
\caption[]{The cross-correlation profile of Gl 57 convolved with our 
theoretical rotational profiles. In order of increasing FWHM: 0, 1.0, 2.0, 3.0,
4.0, 5.0, 7.0, and 10.0 $km\ s^{-1}$. One can see in this figure that the 
cross-correlation profile broadened at 1.0 $km s^{-1}$ is clearly 
distinguishable from the unbroadened cross-correlation profile of Gl 57 (solid 
line).}
\end{figure*}

\begin{figure*} 
\vspace{-1.5cm}
\hspace{-2.5cm}
\includegraphics[width=16cm,angle=-90]{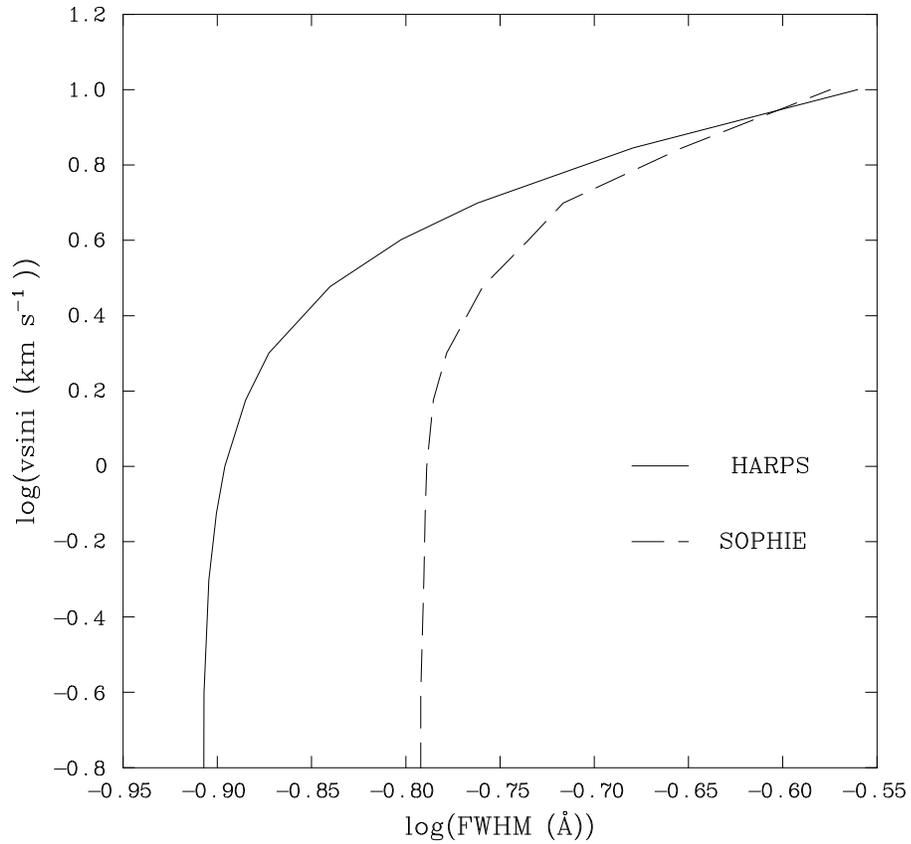}
\vspace{-0.5cm}
\caption[]{The true $v\sin i$ (in $km\ s^{-1}$) as a function of the measured 
FWHM (in \AA ) of the cross-correlation peak. These curves are used to compute 
a star's $v\sin i$ from HARPS (continuous line) and SOPHIE (dashed line) 
measurements of FWHM. The typical uncertainty on these curves is $\pm$0.005\AA\
 on the X-axis.}
\end{figure*}

\begin{figure*} 
\vspace{-0.5cm}
\hspace{-4.5cm}
\includegraphics[width=8cm,angle=-90]{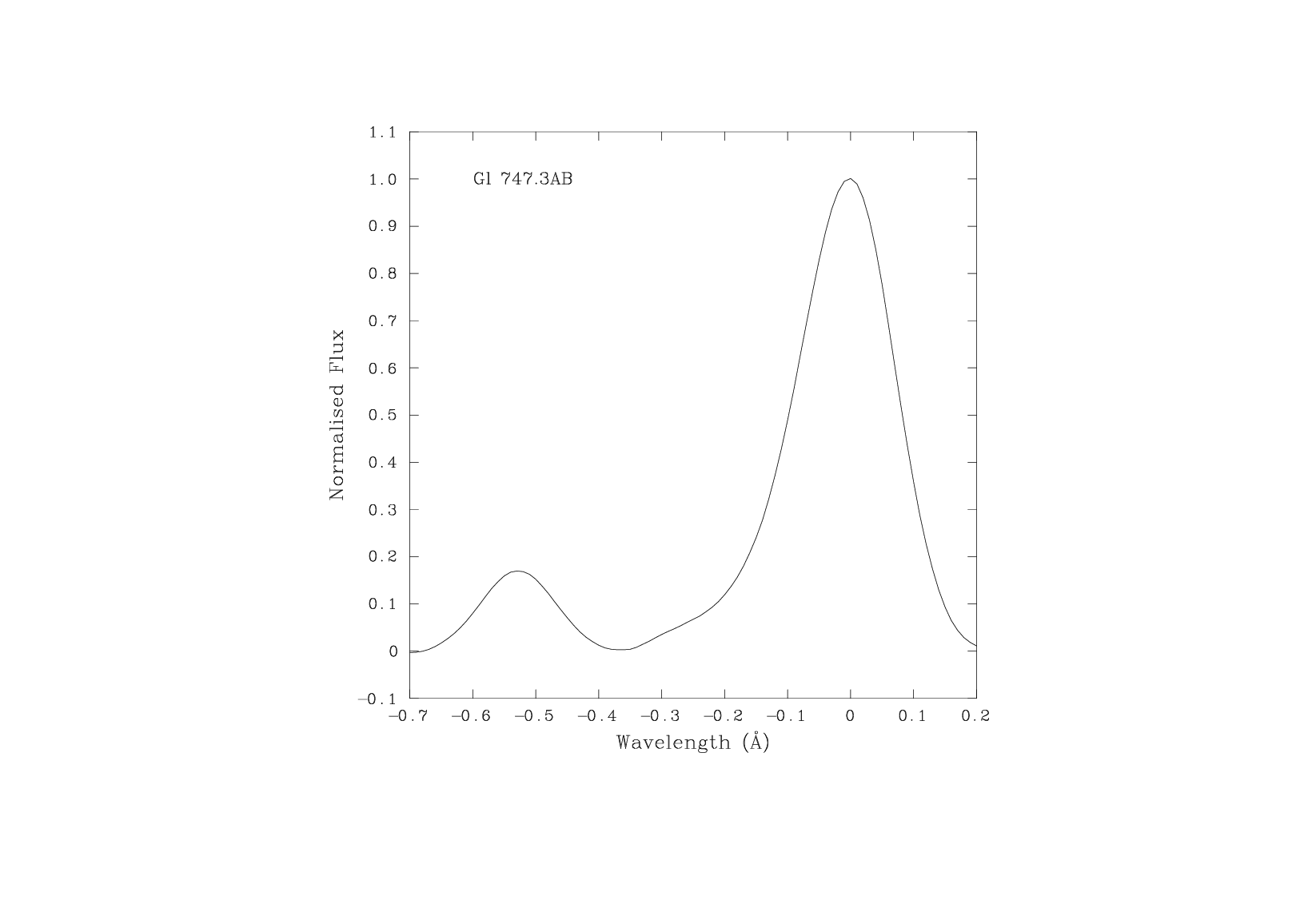}
\hspace{-4.5cm}
\includegraphics[width=8cm,angle=-90]{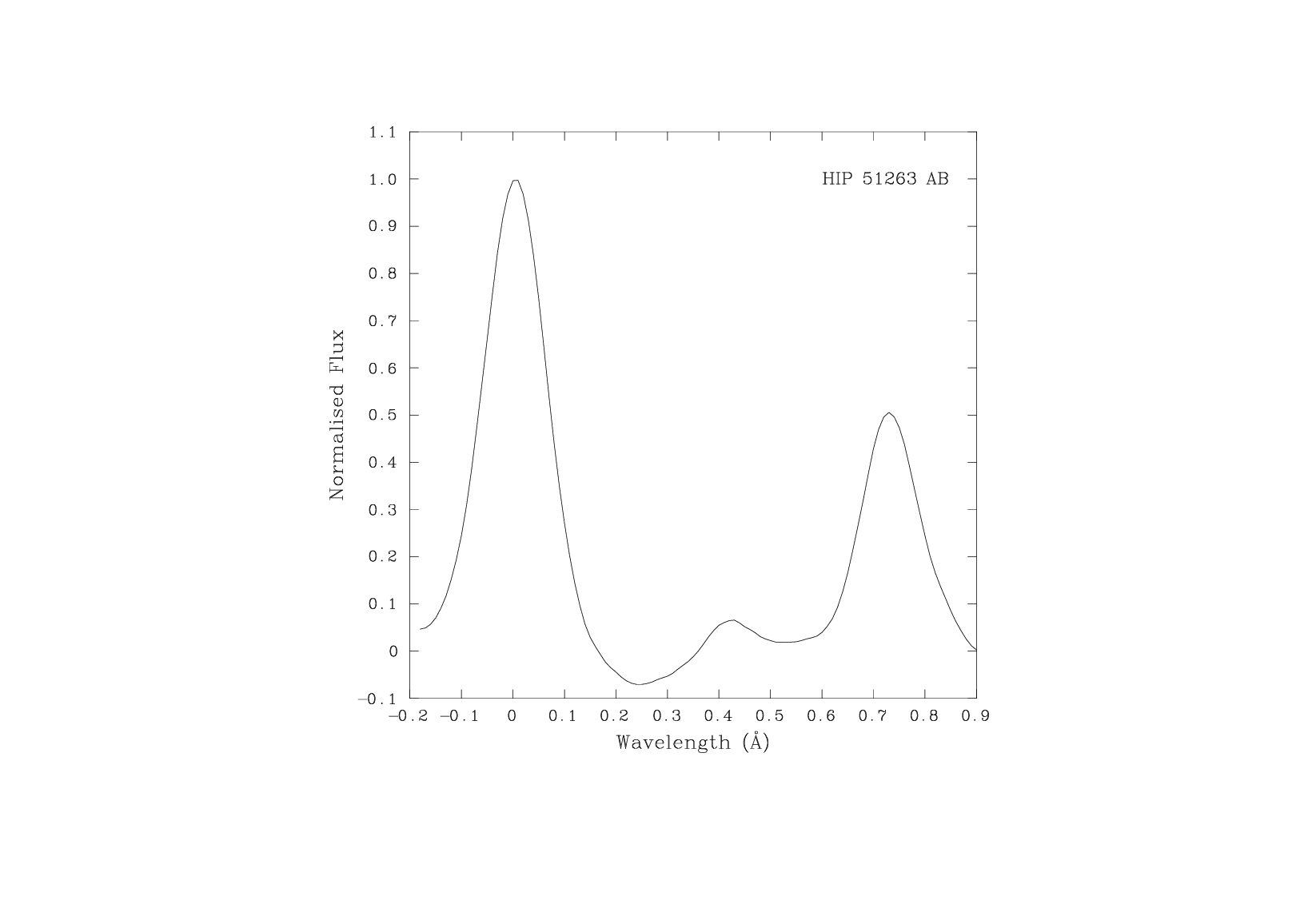}
\includegraphics[width=8cm,angle=-90]{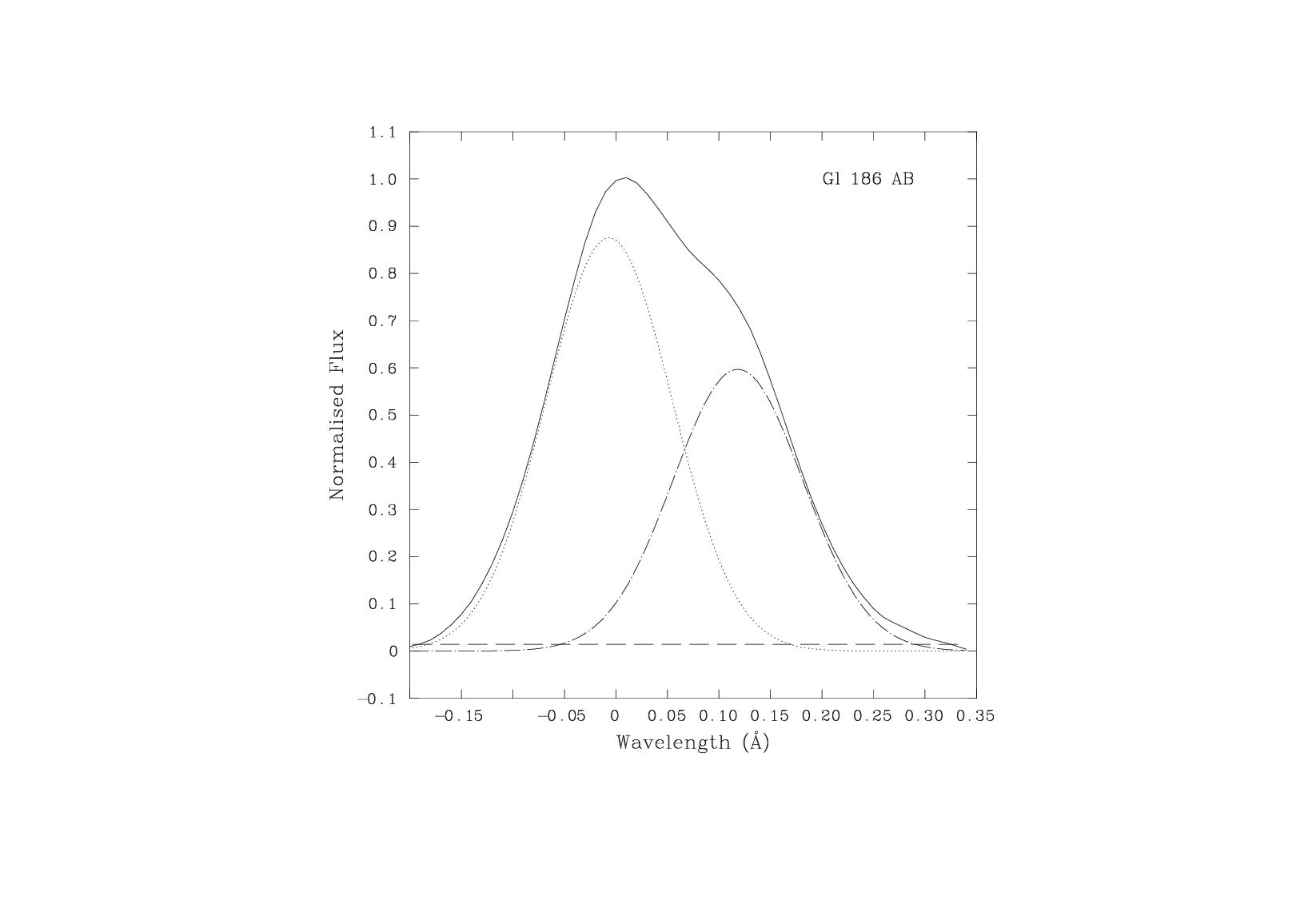}
\vspace{-0.5cm}
\caption[]{Cross-correlation functions for three spectroscopic binaries in our 
late-K dwarf sample: Gl 186AB, Gl 747.3AB and HIP 51263AB. For two of these 
binaries, the two components are well separated and we do not need to use 
multi-Gaussian fits in our analysis. Moreover, in these two cases, the 
secondary is an M dwarf of later spectral type, and as a result is 
considerably fainter than the primary. In the case of Gl 186AB we analyzed 
this system using multi-Gaussian fit to the cross-correlation function. We 
illustrate the best fit (superposed) as well as the profiles of the 
individual components.}
\end{figure*}

\begin{figure*} 
\vspace{-0.5cm}
\hspace{-4.5cm}
\includegraphics[width=8cm,angle=-90]{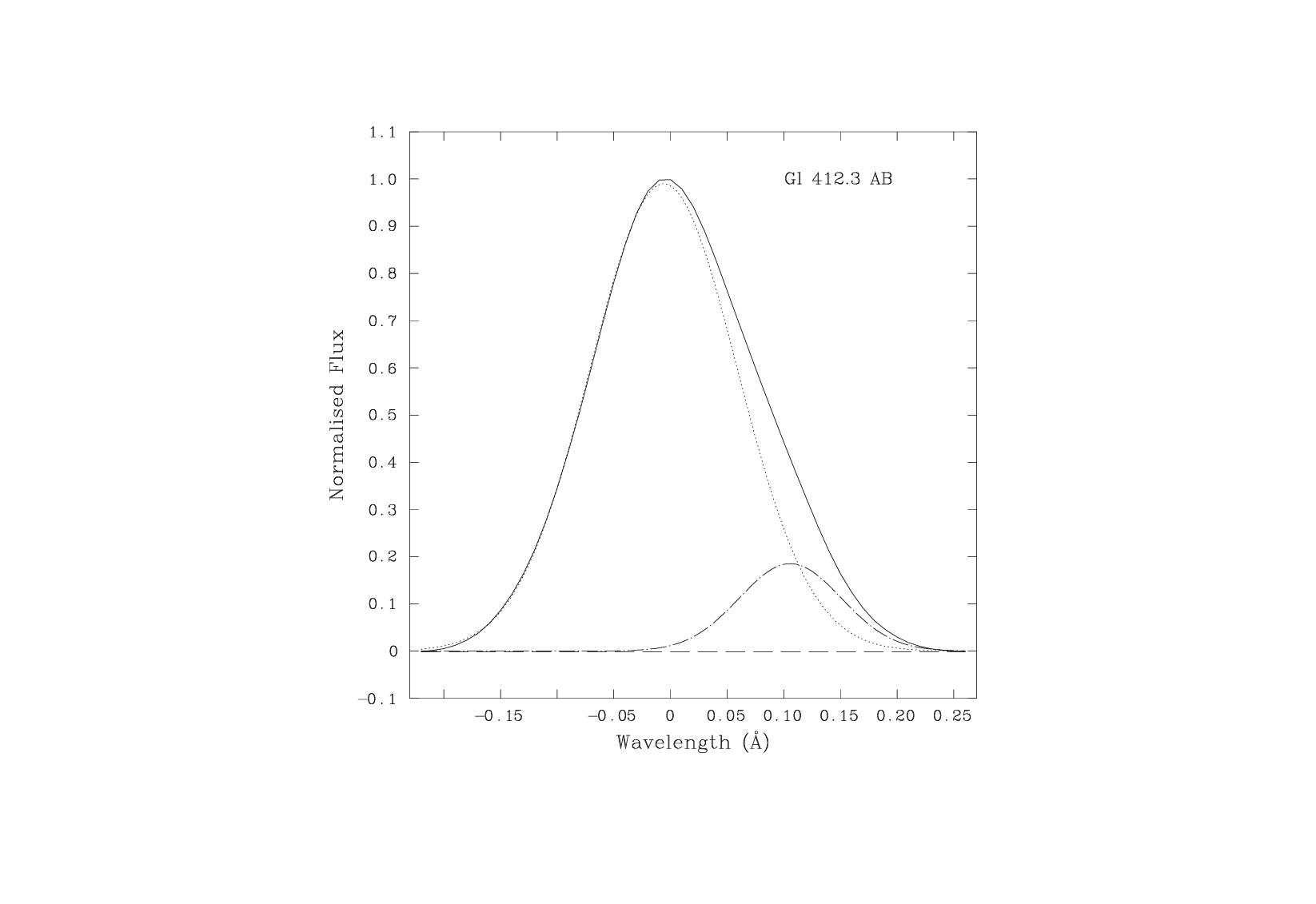}
\hspace{-4.5cm}
\includegraphics[width=8cm,angle=-90]{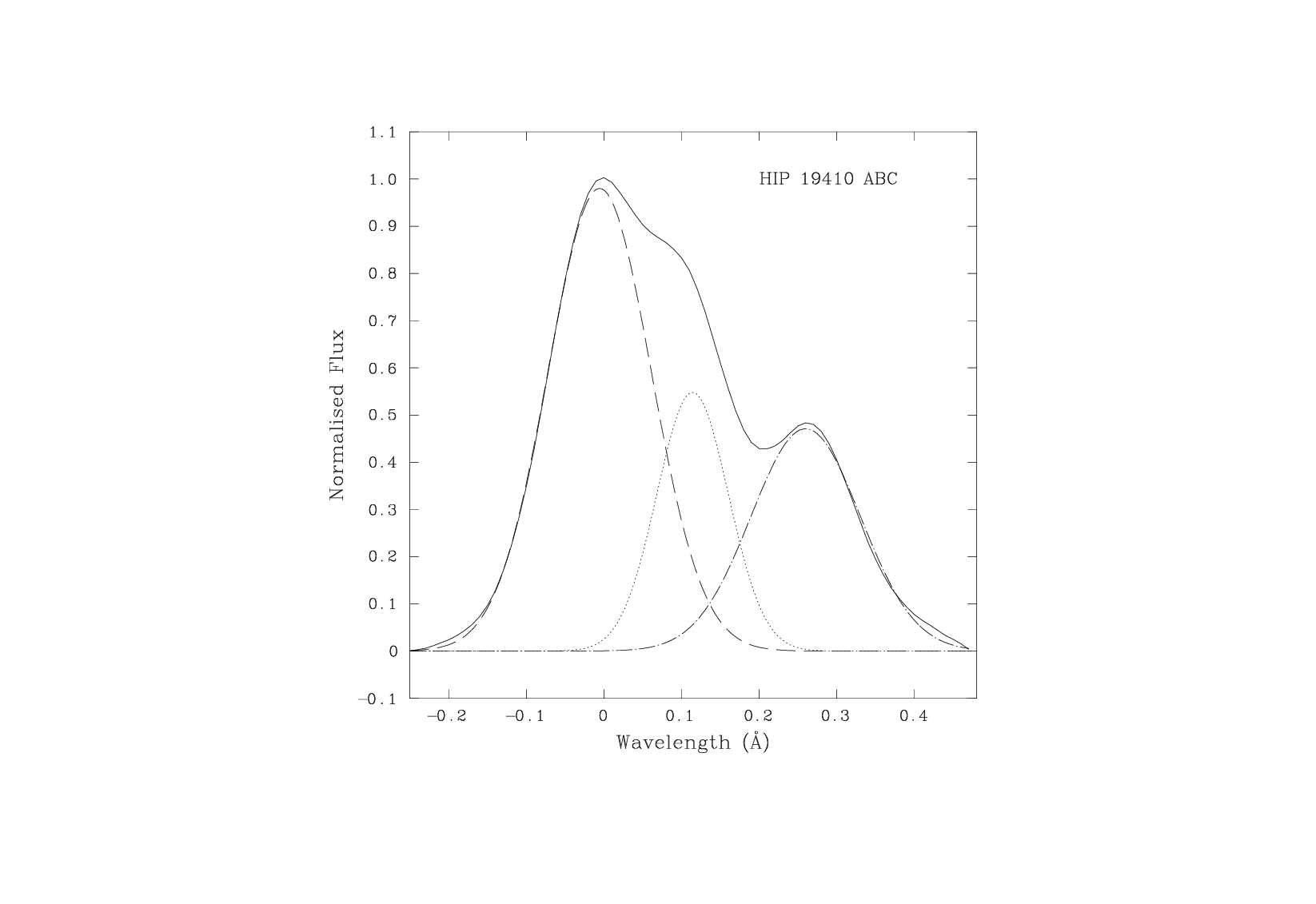}
\includegraphics[width=8cm,angle=-90]{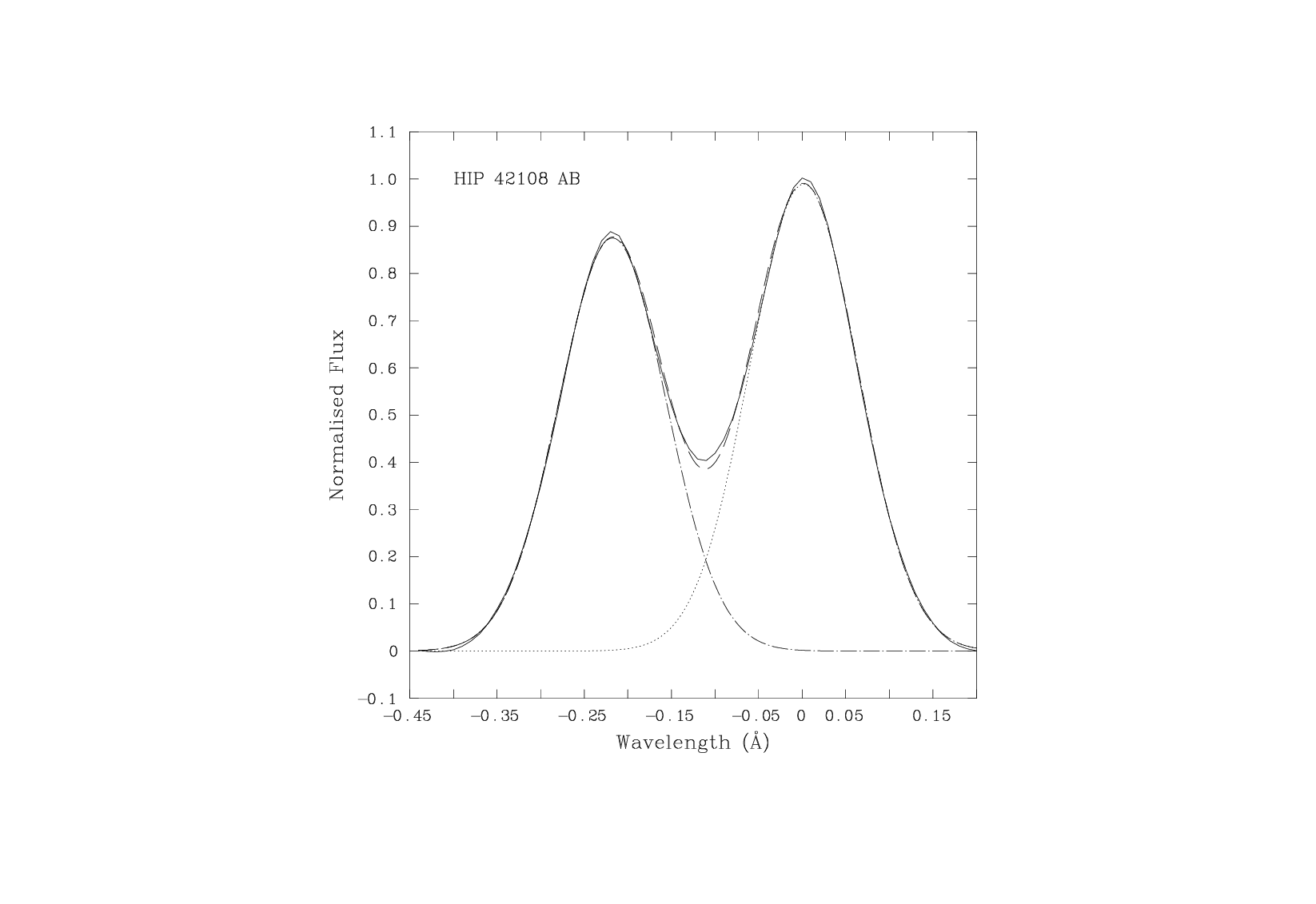}
\vspace{-0.5cm}
\caption[]{Cross-correlation functions for two spectroscopic binaries and one 
triple system in our late-K dwarf sample: Gl 412.3AB, HIP 19410ABC, and HIP 
42108AB. We analyzed these systems using multi-Gaussian fits to the 
cross-correlation functions. In each figure we illustrate the best fit 
(superposed) as well as the profiles of the individual components.}
\end{figure*}

We show in Fig.~7 the cross-correlation profiles obtained with HARPS 
observations for 5 late-K dwarfs in our sample:  Gl 57 (0 $km s^{-1}$), GJ 
9299 (1.78 $km\ s^{-1}$), GJ 9714 (2.51 $km\ s^{-1}$), Gl 425B (3.35 $km\ 
s^{-1}$) and Gl 517 (9.77 $km\ s^{-1}$) (from narrowest to broadest). The 
FWHM of the 
peak for any particular star (in units of \AA ) is used to determine the 
value of $v\sin i$ for that star. In order to measure the FWHM of the 
cross-correlation functions, we measured the background level on each side of 
the cross-correlation functions: then we subtracted the linear interpolation 
of this background, and normalized the functions. This ensures that we isolate 
only the cross-correlation profiles in our results, with little influence of 
the background. We then measured directly the FWHM from the profiles. We note 
that we have very high S/N ratios (typically severall hundreds) and a good 
sampling for the cross-correlation functions (0.01\AA).

As in our previous papers, we take into account limb-darkening effects before 
we can derive significant $v\sin i$ measures for our rotators. We computed 
synthetic rotational profiles that include limb darkening. The synthetic  
profiles were computed according to the formulation of Gray (1988) with the 
center to limb darkening after Claret (2000). We computed synthetic profiles 
for the following range of $v\sin i$ values (in $km\ s^{-1}$): 0.05, 0.25, 
0.50, 0.75, 1.00, 1.50, 2.00, 3.00, 4.0, 5.0, 7.0, and 10.0. 

We convolve our cross-correlation functions of our template stars (Gl 57 for 
HARPS and Gl 728 for SOPHIE) with our rotational profiles. The FWHM of Gl 57 
is 0.1236\AA\ (6.40 $km s^{-1}$) and for Gl 728 it is 0.1611\AA\ (8.35 
$km s^{-1}$). Some of the cross-correlation profiles of Gl 57 convolved with 
our theoretical rotational profiles are shown in Fig.~8 in order of increasing 
FWHM: 0, 1.0, 2.0, 3.0,4.0, 5.0, 7.0, and 10.0 $km s^{-1}$. One can see in 
this figure that the cross-correlation profile broadened at 1.0 $km s^{-1}$ is 
clearly distinguishable from the unbroadened cross-correlation profile of 
Gl 57 (solid line).

We measured the FWHM of the rotational profiles and show these measurements 
as a function of the theoretical $v\sin i$ in Fig.~9 for HARPS and SOPHIE. We 
use this diagram to derive $v\sin i$ values from the direct FWHM measures. 
In general, the correct $v\sin i$ values are smaller than the corresponding 
FWHM values assuming a Gaussian profile.

In Table~7, we list the values we have derived for $v\sin i$ for our late K 
dwarfs. Rotation was detected in most of our sample (92 stars), with $v\sin i$ 
typically in the range 1.0-3.5 $km\ s^{-1}$.

With HARPS, we also observed a few stars of particular interest; five 
spectroscopic binaries (Gl~186AB, Gl~412.3AB, Gl~747.3AB, HIP~42108AB, 
HIP~51263AB) and one spectroscopic triple system (HIP~19410ABC). For two of 
these stars, the components are well separated (Gl~747.3AB and HIP~51263AB). 
We show their cross-correlation profiles in Fig.~10. One can see that in both 
cases the secondary peak is much weaker than the primary peak. This indicates 
that the secondary has a later spectral type than that of the primary. 
For these two spectroscopic binaries, we proceeded as in the case of single 
stars, i.e., we measured the FWHM directly from the correlation profiles. 
For the other stars, the two or three components are blended. In these cases 
we applied multi-Gaussian fits to the cross-correlation profiles. We show 
these profiles in Fig.~11 together with the fits and the profiles of each 
component.

In the case of Gl~186AB, the secondary is significantly weaker than the 
primary. It is therefore most likely not a late K dwarf. The same applies to 
the secondary and tertiary components of HIP~19410. On the contrary, the 
primary and secondary components of HIP~42108AB have almost the same intensity.
 In this case, we assume that both components are late K dwarfs. Note, the 
relatively good fits that we obtain with multi-Gaussians. We give the FWHM of 
the cross-correlation peaks for all these stars in Table~7 together with the 
uncertainties on the fits. 

We would like to emphasize that a few of the active stars in our dK6 stellar 
sample are candidate young stars that may not yet have contracted to the Main-
Sequence (MS). A reliable way to idendify the (few) Pre-Main-Sequence (PMS) 
stars in our samples is the stellar radius: the fact is, that PMS stars that 
have not yet contracted to the MS have abnormally large radii. We identified 
three such stars in our dK6 sample: GJ 1177A, GJ 182 and GJ 425B as possible 
PMS stars (see Table~1). Of these, Gl 425B is a rather low activity star. As 
such, the abnormally large radius for this star is probably due to binarity. 
Nevertheless, we shall find (in Paper~II) that these stars {\em do correlate 
well with the MS stars in the Rotation-Activity Correlation} (RAC). In our M2 
sample (see Table~3), we have identified also two PMS stars: GJ 1264 and 
GJ 803. But again, as we shall see in Paper~II, these PMS stars do correlate 
very well with the other MS stars in the context of the RAC. In our M3 sample, 
GJ 277A is a possible PMS star (Table~4). In our M4 sample, GJ 2069A, GJ 3322, 
GJ 4185A, GJ 4338B, GJ 669A, and GJ 812A are possible PMS stars according to 
their radii (Table~5). {\em We emphasize that all these stars do not rotate 
especially fast. There are many MS stars that rotate faster.} These PMS stars 
are supposed actually to spin up as they contract to the MS, and young MS 
stars are supposed to be the fastest rotators (e.g. Barnes 2003). That is what 
we actually observe in our samples of stars: Young MS stars are the fastest 
rotators (e.g. among our M4 sample; GJ 3631, GJ 3789, GJ 4020B, GJ 4338B, 
GJ 431, GJ 630.1, GJ 791.2A) with $v\sin i$ in excess of 15 $km\ s^{-1}$ and 
up to 56 $km\ s^{-1}$.

All our RACs will turn out (in Paper II) to demonstrate that PMS stars do not 
depart significantly from the overall RAC's of the MS stars with the same 
spectral sub-type. Numerous previous studies also found similar results: e.g. 
Mamajek \& Hillebrand (2008), Browning et al. (2010), Christian et al. (2011), 
West et al. (2015). This finding is rather intriguing since PMS stars may have 
internal structures which are different from MS stars: some PMS stars may even 
be fully convective. More investigation is required to confirm this result, 
but so far, in our samples, we found no evidence that PMS stars obey 
significantly different RACs from those of MS stars (see Paper~II).

\section{Sources of uncertainty in the measures}

For dM2 stars (Paper XIV), we showed that, by using four different 
spectrographs and the results from other authors, it was possible to attain a 
detection limit of 1~$km\ s^{-1}$ in $v\sin i$, with a precision of about 
0.3~$km\ s^{-1}$. This empirical analysis includes all sources of 
uncertainties on the measures. 

In the present study, as was the case for our dM2 stars, our cross-correlation 
profiles have a very high S/N ratio (several hundreds, see Fig.~7). This is 
due to the fact that many spectral lines enter into our calculation of the 
cross-correlations. It is therefore in theory possible to measure subtle 
differences between different cross-correlation profiles of only a few percent.
 However, as in the case of dM2 stars (mentioned above) as well in the case of 
dM3 stars (see HM), the greatest uncertainty arises from the background of the 
cross-correlation functions. There are secondary peaks in the cross-correlation
 function backgrounds and this represents a problem in defining the zero level.
 In fact, these secondary features repeat in all of our cross-correlations 
because all of our stars have quite similar spectral types. For dK6 stars, we 
found that the noise in the background is a bit greater than that of dM2 and 
especially dM3 stars. This is due to the nature of the spectral range we use 
for the cross-correlations. In dM3 stars, there are hundreds of narrow 
spectral lines present in the spectrum, and these lines are evenly distributed 
across the selected wavelength range. However, for dK6 stars, the spectral 
lines are stronger, and there are fewer of them within our selected wavelength 
range. This results in a larger background noise when we calculate the 
cross-correlation functions.

In order to assess the noise in the background from one star to another, we 
averaged the backgrounds, after normalisation of the cross-correlation peaks, 
of four stars: Gl 45, Gl~57, Gl~162.2 and Gl~425B. We then subtracted this 
average background to the raw cross-correlation profiles: the resulting 
profiles and remaining background fluctuations are shown in Fig.~12. One can 
notice some differences in the broadenings of the cross-correlation functions, 
Gl~57, Gl~45, Gl~162.2, Gl~425B, but also some variations in the 
cross-correlation backgrounds. This gives us a direct estimate of the typical 
background ``noise". This noise (i.e. the secondary peaks) has 
an amplitude of about $\pm 3\sigma =$3\% of the main peak (Fig. 12). The noise 
level in the dK6 analysis implies an uncertainty of $3\sigma$=0.0028\AA\ in 
the FWHM, i.e., $3\sigma = 0.40 km\ s^{-1}$ on $v\sin i$ for HARPS spectra for 
$v\sin i = 2 km\ s^{-1}$. For SOPHIE spectra we have an uncertainty of about 
$3\sigma$=0.0034\AA . However, these uncertainties on the FWHM yield a 
variable uncertainty on $v\sin i$ depending on where the data fall relative to 
the curves in Fig.~9. Indeed, because the curves in Fig.~9 are significantly 
steeper at small FWHM, the uncertainty in $v\sin i$ increases, the smaller the 
value of $v\sin i$. The curves in Fig.~9 show that $v\sin i$ is not detected 
within $\pm 3\sigma$ below 0.88~$km\ s^{-1}$ for HARPS and 1.67~$km\ s^{-1}$ 
for SOPHIE.

\begin{figure*} 
\vspace{-1.5cm}
\hspace{-4.5cm}
\includegraphics[width=16cm,angle=-90]{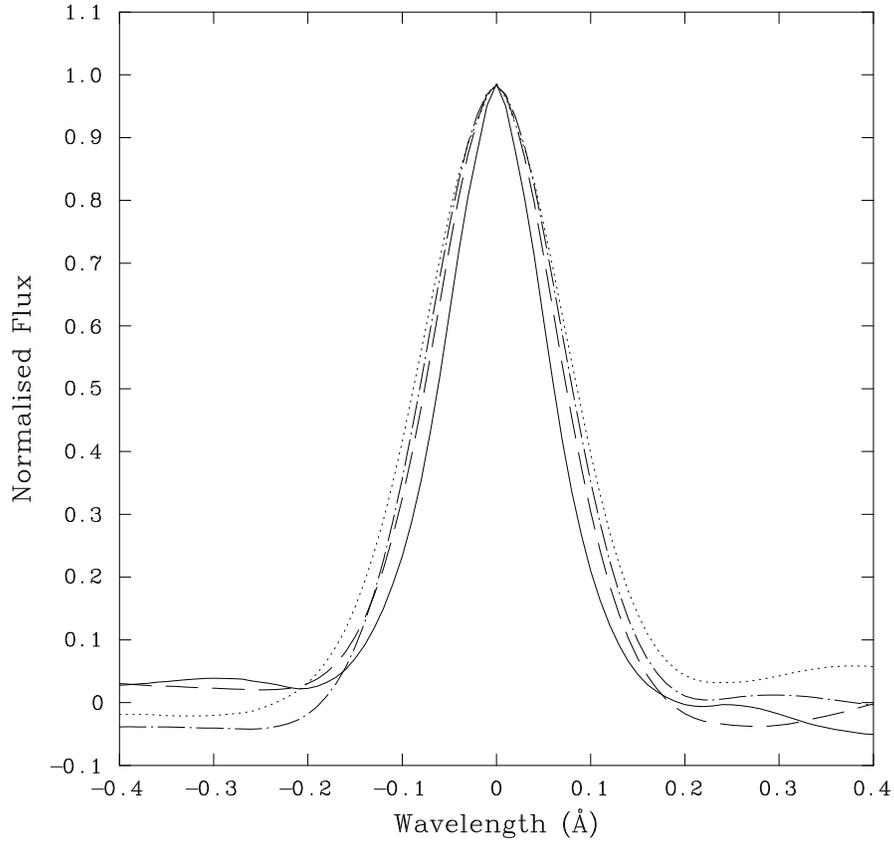}
\vspace{-0.5cm}
\caption[]{The cross-correlation profiles and the variations in the background 
continuum for four late-K dwarfs: Gl~57 (solid line), Gl~45 (dashed line), 
Gl~162.2 (dot-dashed line) and Gl~425B (dotted line). This diagram illustrates 
what are the variations in the background continuum from one star to another, 
and yields an estimate of the error on the FWHM measurements of the 
cross-correlation peaks.} 
\end{figure*}

In order to further assess the precision of our measurements from our 
cross-correlations, we plot in Fig.~13 the values of $v\sin i$ obtained from 
the PCA-based inversion of the stellar parameters (Sect.~3.1) versus the 
values of $v\sin i$ measured from our cross-correlations: the agreement 
between the results from these two independent methods is reasonable if one 
considers the relatively high uncertainty from the PCA-based inversion method 
(1.5 $km\ s^{-1}$). We find that the mean of the differences between the two 
sets of measures is only 0.633 $km\ s^{-1}$. The standard deviation of the 
differences in this set of measurements compared to the one-to-one 
relationship (see Fig.~13) is $\sigma = 0.421 km\ s^{-1}$. This means that the 
typical $3\sigma$ uncertainty for these measures is $1.263 km\ s^{-1}$.
This is better than the expected uncertainty of the PCA-based inversion method 
(1.5 $km\ s^{-1}$, see Sect.~3.1).

\begin{figure*} 
\vspace{-1.5cm}
\hspace{-4.5cm}
\includegraphics[width=16cm,angle=-90]{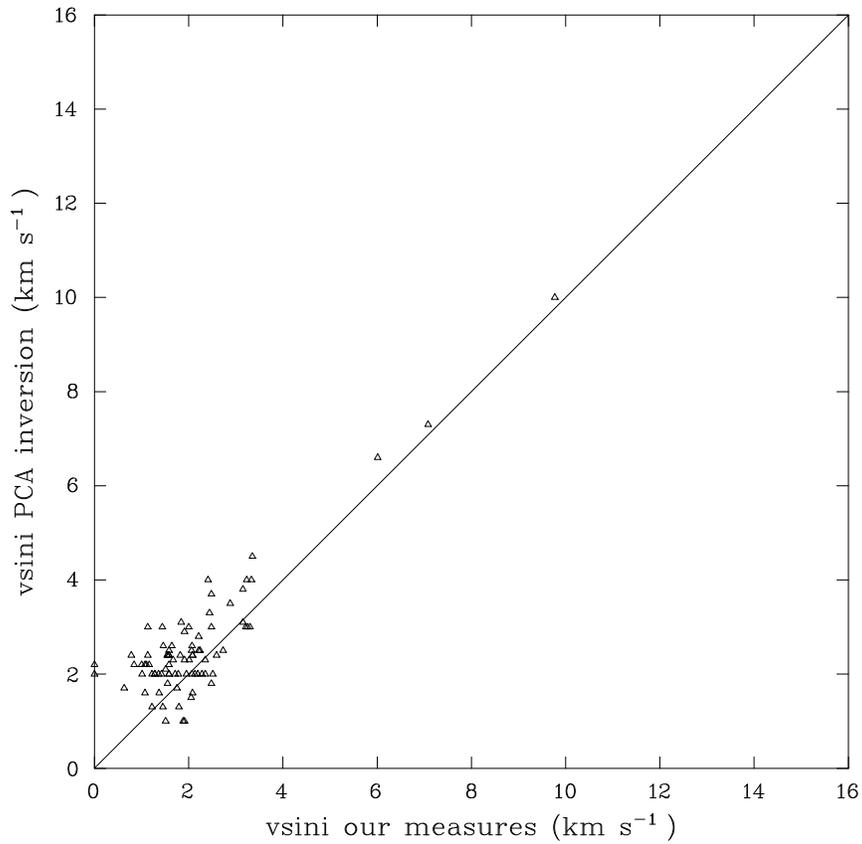}
\vspace{-0.5cm}
\caption[]{Comparison between our cross-correlation results for $v\sin i$ 
(x-axis) and results from our PCA-based inversion of the stellar parameters 
(Sect.~3.1) (y-axis). The solid line indicates the one-to-one relation. 
Differences between the two independent methods do not exceed 0.63~$km\ s^{-1}$
 in average.} 
\end{figure*}

We searched the literature for other $v\sin i$ or rotational period $P$ 
measurements for our K6, M2, M3 and M4 targets (we used the tutorial developed 
by Paletou \& Zolotukhin 2014\footnote{http://www.astropy.org/}). We found 
several measurements of $v\sin i$ (Vogt et al. 1983, Marcy \& Chen 1992, 
Tokovinin 1992, Favata et al. 1995, Delfosse et al. 1998, Mohanty \& Basri 
2003, Nordstrom et al. 2004, Glebocki \& Gnacinski 2005, Torres et al. 2006, 
Schroder et al. 2009, Jenkins et al. 2009, Mart\'inez-Arn\'aiz et al. 2010) 
and $P$ (Pizzolato et al. 2003, Kiraga \& Stepie\'n 2007, Irwin et al. 2011). 
In cases where a rotation period $P$ was reported in the above papers, we 
derived the $v\sin i$ from $P$ assuming $\sin i=1$. 

We show in Fig.~14 our results for $v\sin i$ versus the results from other 
spectrographs (e.g. SOPHIE as a function of HARPS [Paper XIV, HM], 19 
measurements in total) and the measurements from other authors (73 
measurements). Filled circles denote results obtained for dK6 stars in this 
paper. The overall agreement between our measures and those of other authors 
is good. The agreement goes down to $v\sin i$ as low as 0.5~$km\ s^{-1}$. 
For results below 1~$km\ s^{-1}$ we have 13 data points. For these points, 
the mean difference between our results and others is found to be 0.37~$km\ 
s^{-1}$. The standard deviation of the differences in this set of measurements 
compared to the one-to-one relationship (see Fig.~14) is $\sigma = 0.27 km\ 
s^{-1}$. This means that the typical $3\sigma$ uncertainty for this sub-set of 
measures is only $0.81 km\ s^{-1}$.

For the slow rotators, i.e. for dK6, dM2, dM3 and dM4 stars (with no evidence 
for emission in H$_{\alpha}$), the mean of the differences between our $v\sin 
i$ results and those reported by other authors is only 0.50~$km\ s^{-1}$. The 
standard deviation of the differences in this set of measurements 
compared to the one-to-one relationship is $\sigma = 0.32 km\ s^{-1}$. This 
means that the typical $3\sigma$ uncertainty for the slow rotators is only 
$0.96 km\ s^{-1}$. These figures are lower than the ones we derived from the 
uncertainties arising from features in the background of the 
cross-correlations (see above and HM). This indicates that the uncertainty on 
our $v\sin i$ measures when using the cross-correlation technique depends 
essentially on the S/N ratio in the cross-correlation peaks, and that the 
effects of the uncertainties on the background continuum have been 
overestimated. Therefore, the uncertainties on $v\sin i$ and $P/\sin i$ 
derived in Table~7 are only safe upper limits. These results confirm that both 
HARPS and SOPHIE are very stable instruments on the long-term: therefore, our  
spectra are well suited for reliable measurements of small rotational 
broadenings in K and M dwarfs.

For the fast rotators, i.e. for dK6e, dM2e, dM3e and dM4e stars (with clear 
evidence for emission in H$_{\alpha}$), the mean of the difference between our 
$v\sin i$ results and those reported by others is 0.99~$km\ s^{-1}$. The 
standard deviation of the differences in this set of measurements compared to 
the one-to-one relationship is $\sigma = 0.60 km\ s^{-1}$. This means that the 
typical $3\sigma$ uncertainty for the fast rotators is about $1.80 km\ s^{-1}$.
 This uncertainty is about twice larger than for the slow rotators. This is 
probably due to the presence of large spots on the surface of these stars. 

The reason why we can attain precisions in $v\sin i$ as good as those 
mentioned above is that we have paid attention to several factors in our 
analysis. (i) The choice of the wavelength interval is crucial in obtaining 
cross-correlation profiles where the S/N is high, and the background noise is 
as low as possible. (ii) The method of subtracting the background noise is 
essential in order to recover with the highest precision the true 
cross-correlation profiles. (iii) The spectra which we selected for analysis 
are mostly high S/N ratio spectra, and were all reduced with an identical 
automated technique. (iv) We use only spectrographs that are very stable in 
radial velocity and that use image scramblers to avoid seeing effects on the 
FWHM of the spectral lines. By combining these factors, we have found that it 
is possible to measure subtle differences between different cross-correlation 
profiles at the level of as little as a few percent. 

\begin{figure*} 
\vspace{-1.5cm}
\hspace{-6.5cm}
\includegraphics[width=20cm,angle=-90]{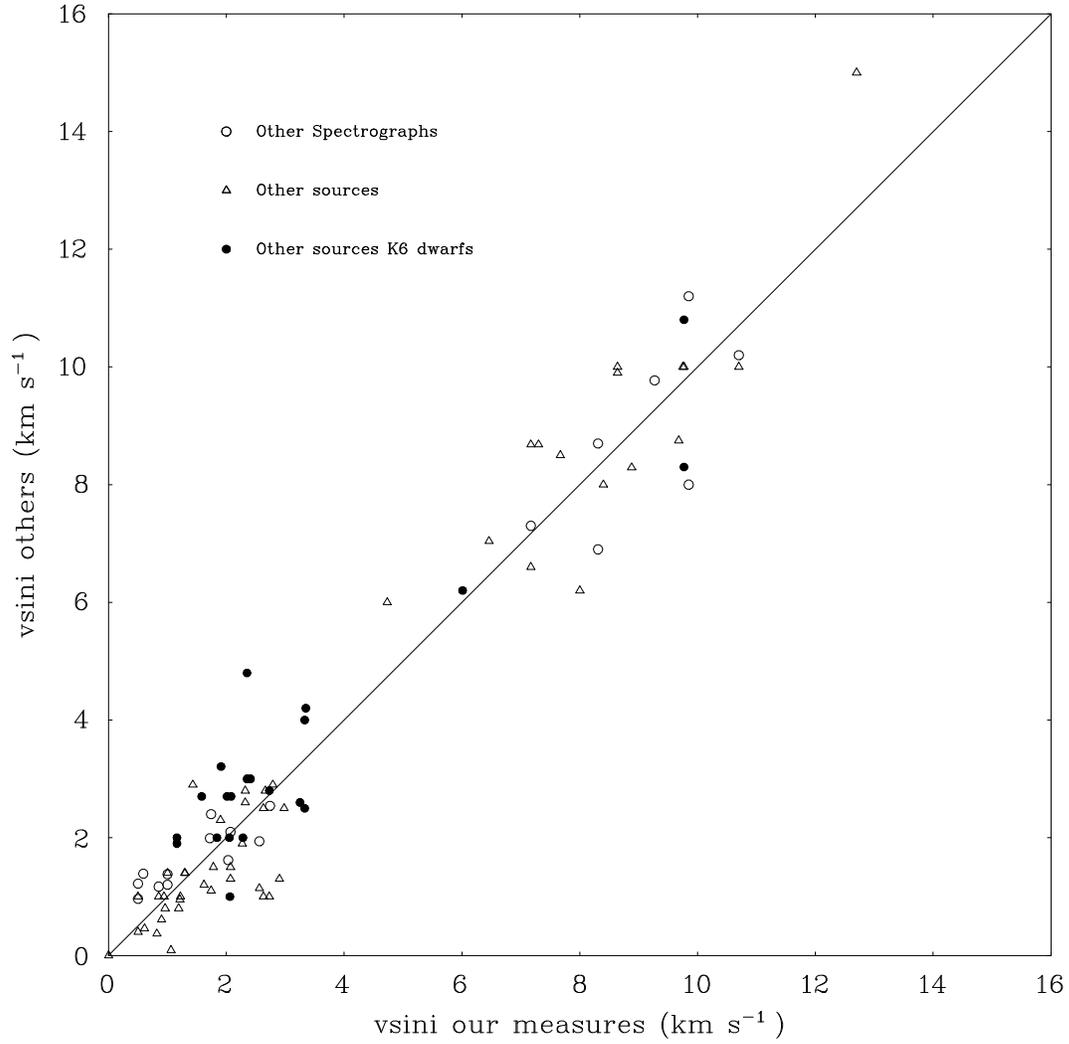}
\vspace{-0.5cm}
\caption[]{Comparison between our results for $v\sin i$ (x-axis) and results 
from other authors (y-axis). We show the K6 dwarf measurements of other 
sources as a function of the present measurements as filled circles. Included 
in the plot are data for all of our samples of M dwarfs (including sub-types 
K6, M2, M3 and M4). The solid line indicates the one-to-one relation. 
Differences between our results and the others do not exceed 0.50~$km\ s^{-1}$ 
in average for the low activity stars.} 
\end{figure*}

\section{The distributions of rotational periods among M and K dwarfs}

We derived the projected rotation periods $P/\sin i$ for all our dK6 targets 
according to the $v\sin i$ results in Table~7 and the radii given in Table~1. 
Values of $P/\sin i$ are given in Table~7 for our dK6 target stars, along with 
their uncertainties.

We show in Fig.~12 the histogram of $P/\sin i$ values for dK6 stars, together 
with the analogous results for dM2 stars (upper panel). Also shown are the 
histograms of $P/\sin i$ values for dM3 and dM4 stars (lower panel). We see 
that the peak of the distribution for dK6 stars lies at a mean projected 
period in the range 13-22 days. We also see that the longest $P/\sin i$ that 
we have detected for dK6 stars is 30-35 days (see Fig.~15): these longest 
periods overlap with the range of upper limits ($\sim$33-66 days) which were 
estimated in Sect.~1 above, given the limitations (0.5-1$km\ s^{-1}$) on 
measuring $v\ sin i$. 

The histogram for dK6 stars in Fig.~15 has a different shape from that of 
dM2 stars, in spite of their relatively close spectral types. In dM2 stars, in 
addition to a main peak at about 11 days, there may be a second (smaller) peak 
at shorter periods (4 days): if these two peaks are really distinct, we could 
refer to them as belonging to slow and fast rotators respectively, with more 
slow rotators than fast. Among the dK6 sample, there is also an analogue of a 
short-period peak at about 3-7 days. In dM2 stars the distribution among the 
slow rotators is compact, extending to periods no larger than 16-20 days, 
whereas in dK6 stars, the distribution is more spread out, extending to 25 
days. The distribution for dM4 stars is not bimodal as in dM2 and dK6 stars. 
For dM4 stars, there is a hint that there are more fast rotators than slow 
rotators. For dM3 stars, our dataset suggests that the distribution is also 
bimodal, with two distinct groups of slow and fast rotators.

\begin{figure*} 
\vspace{-0.5cm}
\hspace{-4.5cm}
\includegraphics[width=12cm,angle=-90]{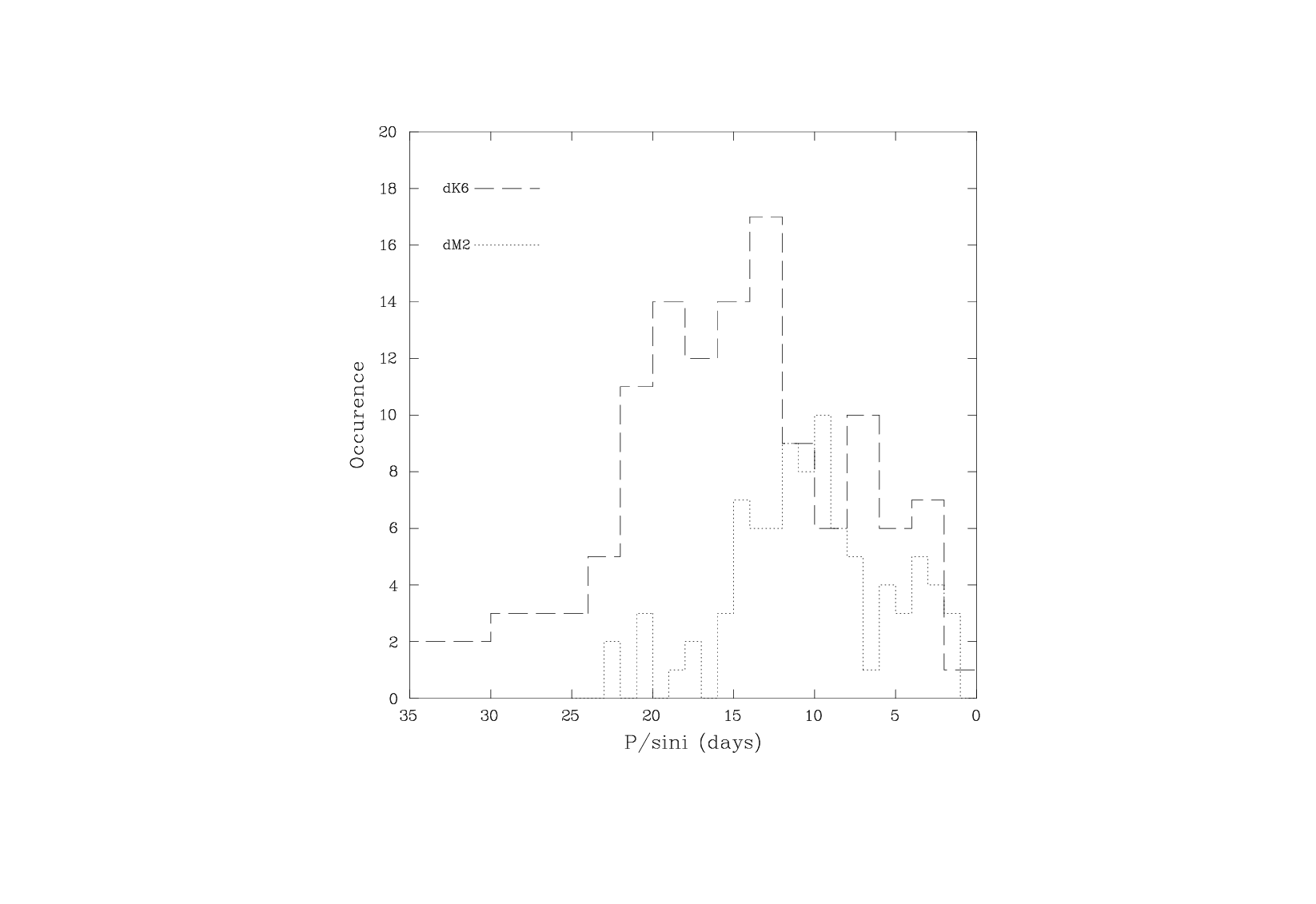}
\hspace{-6.5cm}
\includegraphics[width=12cm,angle=-90]{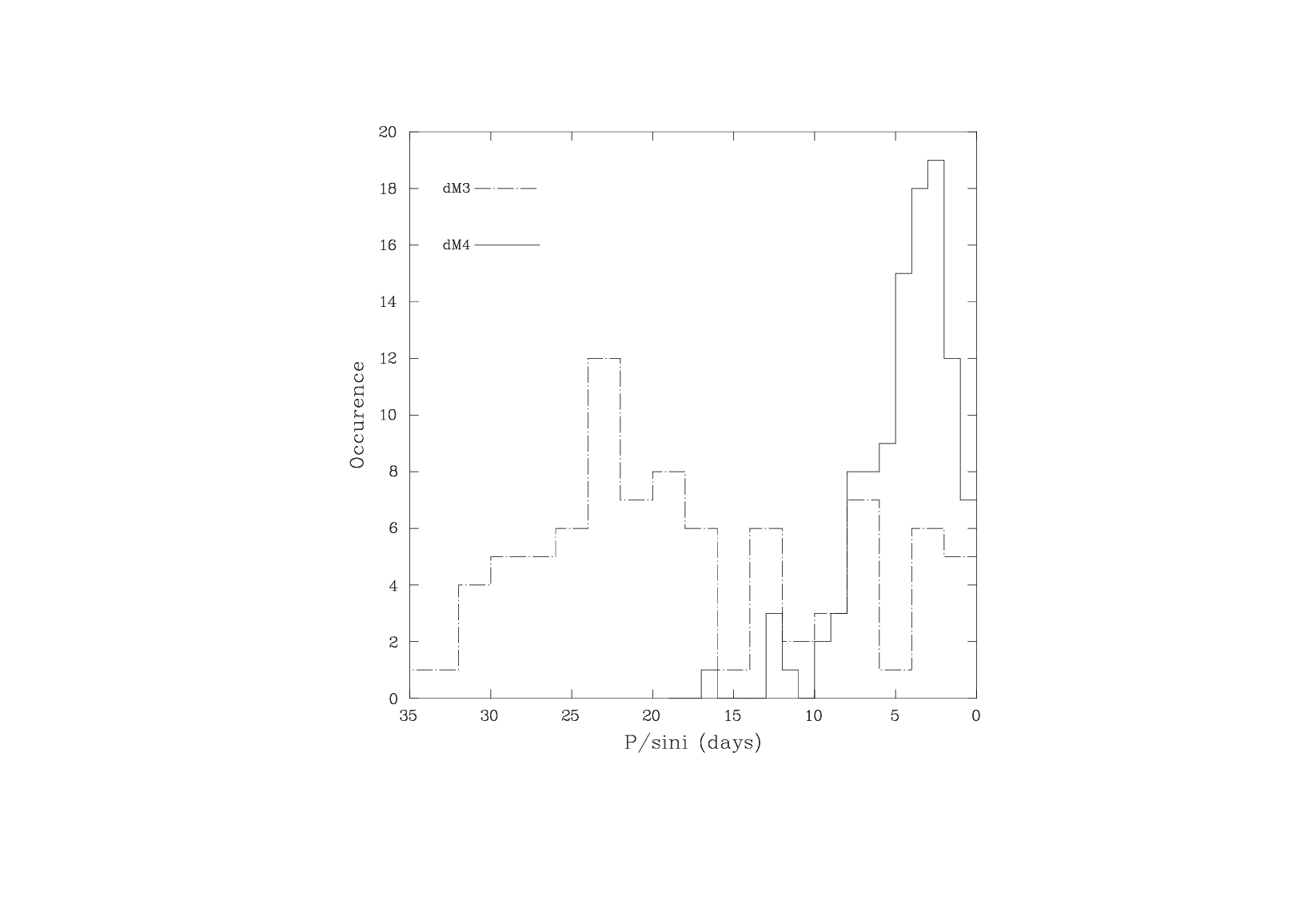}
\vspace{-0.5cm}
\caption[]{Histograms of stars as a function of $P/\sin i$ for dK6 stars, dM2 
stars, dM3 stars and dM4 stars. The peak of the distribution shifts from 
(about) 17 days for dK6 stars, to 10 days for dM2 stars, and to 3 days for 
dM4 stars.}
\end{figure*}

The principal conclusion which we draw from the two panels in Fig.~15 is 
that there appears to be a global shift of the distributions from dK6 stars to 
dM4 stars. Specifically, mid-M dwarfs rotate in general faster than early M 
dwarfs, a fact already well established among M dwarfs. The differences 
between the 4 histograms in Fig.~15 indicate the value of selecting stars in 
narrowly restricted spectral sub-types: this selection suggests that different 
sub-types among the M dwarfs may exhibit distinctive rotational properties. 
Finally, we note that the mean period of the dK6 star histogram (17 days) is 
intermediate between the mean period of the dM2 star histogram (11 days) and 
that of dK4 stars ($\approx$ 35 days, Paper XVI). This is an indication of the 
overall trend towards shorter rotation periods as we go from dK4 to dM4 (see 
HM).

Formally, we have calculated a mean value and a median value for each of the 4 
histograms shown in Fig.~15. For dK6 stars, we find a mean of 15.8 days, and a 
median of 16.0 days. For dM2 stars, the mean and median are 10.8 days and 11.5 
days respectively. For dM3 stars, the mean and median are 18.0 days and 20.5 
days respectively. For dM4 stars, the mean and median are 4.9 days and 5.0 
days respectively. Thus, whether we consider the mean period or the median 
period, we can state the following: as we go from dK6 to dM2 to dM4, there is 
a trend towards shorter periods. However, there is an exception to this trend 
at spectral sub-type dM3: there, the mean/median period does not fit the 
trend, but is definitely longer than the dK6-dM2-dM4 trend would predict. In 
fact, the dM3 mean/median period is actually longer than the results for dK6. 
Thus, as already suggested by the smaller data set used by HM, the dM3 stars 
in our sample have markedly slower rotations than the trend between dK6 and 
dM4. 

In HM, histograms were derived for $P/\sin i$ values among dM2 stars (85 
targets), dM3 stars (80 targets), and dM4 stars (21 targets) (see Fig.~9 in 
HM). In the present paper, we have obtained a fourth histogram, for dK6 stars. 
Even a casual inspection shows that there are differences between the 
histograms (see Fig.~15). The question is: can a random distribution of 
inclination angles produce the 4 distributions, assuming the actual period 
distribution is the same for all three samples? 

To address this, we have applied the Kolmogorov-Smirnov (K-S) two-sided test 
to each histogram in order to test the null hypothesis that the stars in any 
one sample are drawn from the same population as the stars in another sample. 
For any pair of samples, we construct the cumulative distributions F(1) and 
F(2) for periods ranging from 44 days to 1 day. Extracting numerical values 
for the K-S statistic $D(1,2)$, we find that when we compare a given 
distribution to the distribution of any other spectral sub-type, the null 
hypothesis (namely, that the distributions are drawn from the same population)
 is rejected at a confidence level which is better than 0.001\%. This suggests 
that the $P/\sin i$ distribution (i.e. the rotational braking mechanism) in 
dM3 stars is controlled by different factors than in dM2 or dM4 stars. The K-S 
analysis also indicates that the dK6 stars have a distribution which differs 
significantly from those of the other dM spectral sub-types.

But in this regard, an anonymous referee has pointed out that histograms can 
suffer from slight distortions as a result of choices related to binning. In 
view of this, the referee suggests that, when comparing the rotational period 
distributions in Fig.~15, it may be more useful to compare smoothed continuous 
distributions, rather the raw histograms. Different possibilities are 
available for smoothing, e.g. kernel density estimates, splines, bezier...  
Here, the software package (gnuplot) which is available to one of the authors 
(DJM)  offers bezier smoothing. Using this, the smoothed rotation distributions
 were used to construct cumulative distributions for all 4 spectral sub-types. 
For each pair of spectral sub-types, the K-S statistic $D(1,2)$ was extracted. 
The maximum value of $D(1,2)$, normalized by a factor which depends on the 
sizes of the two distributions which are being compared, leads to a 
coefficient c(${\alpha}$) which was found to have the following numerical 
values for various sample pairs: 5.067 (K6 versus M4), 4.623 (M3 versus M4), 
3.602 (M2 versus M4), 2.901 (M2 versus M3), 2.201 (K6 versus M2), and 1.382 
(K6 versus M3). Statistically speaking, values of c(${\alpha}$) in excess of 
1.95 correspond to a probability $\alpha$ of less than 0.001 that a pair of 
distributions has been drawn from the same parent population. The above 
numerical results indicate that the rotational distributions of K6 and M3 
stars (for which c(${\alpha}$) = 1.382) {\it might} originate from the same 
population. But in the case of all the other pairs (K6,M2), (K6,M4), (M2,M3), 
(M2,M4), and (M3,M4), the bezier smoothed distributions have a chance of less 
than 0.1$\%$ of being drawn from the same population. This suggests that the 
$P/\sin i$ distribution (which is determined by the rotational braking 
mechanism) in dM3 stars is controlled by different factors from those which 
operate in dM2 or dM4 stars, but that dK6 stars {\it might} have a similar 
braking mechanism to that in dM3 stars.

\section{Conclusion}

In this paper, we have derived numerical values of various parameters of K and 
M dwarfs in preparation for a follow-up study of RAC's (in a future Paper II). 
For M dwarfs, values of $T_{eff}$ have been determined from the (R-I)$_C$ 
color using the empirical calibration of Mann et al. (2015). For late-K 
dwarfs, values of $T_{eff}$ have been determined from the (R-I)$_C$ color 
using the calibration of Kenyon \& Hartmann (1995). These determinations have 
led to values of $T_{eff}$ for a total of 612 stars with spectral sub-types 
dK4, dK6, dM2, dM3 and dM4. In an independent approach, we have also determined
 several stellar parameters ($T_{eff}$, $log(g)$ and [M/H]) using the PCA-based
 inversion technique of Paletou et al. (2015a) for a sample of 105 late-K 
dwarfs. We also compile all effective temperatures from the literature for 
this sample of late-K dwarfs. Together with our determinations of $T_{eff}$ 
from the (R-I)$_C$ color, this allows us to obtain reliable values of $T_{eff}$
 for our sample of late-K dwarfs. With values of $T_{eff}$ available, we then 
derived the stellar radii for our sample of K-M dwarfs, based on parallax 
measurements from Hipparcos. We find that our radii agree well with the 
interferometric radii reported by Boyajian et al. (2012): in the 
$T_{eff}$-radius plane, we find that our estimates of radii are within 0.19\%, 
6.3\%, 4.3\% and 8.2\% of the Boyajian results for our late-K, dM2, dM3 and 
dM4 stellar samples respectively.

We also propose empirical relationships between radius and metallicity for 
dK6, dM3 and dM4 stars, similar to that reported for dM2 stars by Houdebine 
(2011b). This allows us to derive metallicity for all our targets, and these 
will be used in Paper II to correct the RAC's for metallicity effects.

We present a cross-correlation analysis of HARPS and SOPHIE spectra of 105 
late-K dwarfs, which has allowed us to detect $v\sin i$ in 92 of these stars. 
In combination with our previous $v\sin i$ measurements in M and K dwarfs 
(Papers VII, XIV, XVI, XVII, HM) and from the compilations of $v\sin i$ 
measurements from the literature for dK6, dM3 and dM4 stars, we now derive 
$P/\sin i$ measures for a sample of 418 K and M dwarfs. This allows us to 
investigate the distributions of $P/\sin i$ for each spectral sub-type 
separately. We find that the $P/\sin i$ distributions are statistically 
distinct from one spectral sub-type to another at a 99.9\% confidence level. 
Therefore, if one wants to investigate certain properties of K and M dwarfs 
which are related to rotation (such as RAC's), it is important to consider the 
different spectral sub-types separately. 

The parameters which we have derived in this paper for stars belonging to 
various spectral sub-types from dK4 to dM4 will be used as input in Paper II 
to construct separate RAC's for each spectral sub-type. 

\section*{acknowledgements}
This research has made use of the SIMBAD database, operated at CDS, Strasbourg,
France. DJM is supported in part by the NASA Space Grant program. This research
 was based on data obtained from the ESO Science Archive Facility and the 
Observatoire de Haute Provence SOPHIE database. This research made use of 
Astropy\footnote{http://www.astropy.org/ and 
http://astroquery.readthedocs.org/en/latest/}, a community-developed core 
Python package for Astronomy (Astropy Collaboration, 2013)). This research has 
made use of the VizieR catalogue access tool, CDS, Strasbourg, France. The 
original description of the VizieR service was published in A\&AS 143, 23

\onecolumn

\tiny
\begin{landscape}
\begin{longtable}{|lcccccccccccc|}
\caption[ ]{Stellar parameters (e.g. $T_{eff}$, $\pi$, Mv, R$_{\star}$) for 
our sample of late-K dwarfs. We also computed [M/H] derived from the 
[M/H]/radius correlation (see text).} \\

\hline Star & V & (R-I)$_{c}$ or (B-V) & Teff & Teff$\pm \sigma$ & Spect. & $\pi$ & Distance & Mv & R$_{\star}$ & R$_{\star}\footnotemark[1]$& [M/H] & [M/H]\footnotemark[2] \\
        &(mag) & (mag) &     (K)       &     (K)       & Type   &     (m'')     & (pc)   &    (mag)      &  (R$_{\odot}$)&(R$_{\odot}$)&(dex)&(dex)\\
        &      &       &               &               &        &               &        &               &               &         &      &       \\
\endfirsthead
\caption{continued.}\\

\hline Star & V &b(R-I)$_{c}$ or (B-V) & Teff & Teff$\pm \sigma$ & Spect. & $\pi$ & Distance & Mv & R$_{\star}$ & R$_{\star}\footnotemark[1]$& [M/H]&[M/H]\footnotemark[2] \\
        &(mag) & (mag) &     (K)       &     (K)       & Type   &     (m'')     & (pc)   &    (mag)      &  (R$_{\odot}$)&(R$_{\odot}$)&(dex)&(dex)\\
        &      &       &               &               &        &               &        &               &               &         &      &       \\
\endhead  
\hline
\endfoot

        &      &       &               &               &        &               &        &               &                 &       &      &       \\
GJ 1056 & 9.259& 0.686 & 4230 $\pm$100 & 4365 $\pm$103 & dK4.9  & 46.06$\pm$1.08&  21.7  &7.576$\pm$0.055& 0.687$\pm$0.061 & 0.660 &  -   &-0.175 \\
GJ 1066 & 8.847& 0.680 & 4250 $\pm$100 & 4397 $\pm$107 & dK4.8  & 56.31$\pm$0.82&  17.8  &7.600$\pm$0.036& 0.667$\pm$0.052 & 0.634 &  -   &-0.215 \\
GJ 1067 & 9.70 & 0.680 & 4250 $\pm$100 & 4385 $\pm$135 & dK4.9  & 38.26$\pm$1.27&  26.1  &7.614$\pm$0.091& 0.665$\pm$0.071 & 0.634 &  -   &-0.215 \\
GJ 1177A& 8.94 & 0.731 & 4130 $\pm$100 & 4260 $\pm$130 &PMS K5.6& 69.01$\pm$1.35&  14.5  &8.135$\pm$0.062& 0.576$\pm$0.055 & 0.569 &  -   &-0.471 \\
GJ 1213 &10.18 & 0.781 & 4030 $\pm$100 & 4155 $\pm$125 & dK6.4  & 39.64$\pm$1.14&  25.2  &8.171$\pm$0.082& 0.624$\pm$0.072 & 0.634 &  -   &-0.236 \\
GJ 1248 &12.99 & 0.790 & 4010 $\pm$100 & 4010 $\pm$100 &sdK7.3  & 80.2 $\pm$4.3 &  12.5  &12.51$\pm$0.13 & 0.090$\pm$0.012 & 0.140 &  -   &   -   \\
GJ 1267 &8.992 & 0.800 & 4000 $\pm$100 & 4107 $\pm$117 & dK6.7  & 73.90$\pm$1.21&  13.5  &8.335$\pm$0.040& 0.606$\pm$0.058 & 0.629 &  -   &-0.225 \\
\hline 
\end{longtable}
\footnotetext[1]{Radius corrected from T$_{eff}$ effects.}
\footnotetext[2]{Computed from the [M/H]-radius relationship.}
\end{landscape}
\normalsize

\begin{longtable}{|lcccc|}
  \caption{Inverted stellar parameters $T_{\rm eff}$, log$g$, [Fe/H] and 
$v{\rm sin}i$ for HARPS and SOPHIE objects, using our Atlas-based PCA method.} \\

\hline Object & $T_{\rm eff}$ [K]& log$g$ [dex] & [Fe/H] [dex] & $v{\rm sin}i$ [${\rm km\,s}^{-1}$]  \\
  \hline
\endfirsthead
\caption{continued.}\\
\hline Object & $T_{\rm eff}$ [K]& log$g$ [dex] & [Fe/H] [dex] & $v{\rm  sin}i$ [${\rm km\,s}^{-1}$]  \\
\hline
\endhead 
\hline
\endfoot

         &       &      &        &       \\
GJ 1056  &  4480 &  4.8 &  -0.20 &   2.4 \\
GJ 1066  &  4500 &  5.0 &  -0.25 &   1.0 \\
GJ 1067  &  4520 &  4.8 &  -0.22 &   2.4 \\
GJ 1267  &  4270 &  5.0 &  -0.50 &   2.5 \\
GJ 1279  &  4450 &  4.9 &  -0.12 &   2.5 \\
GJ 3072  &  4110 &  4.8 &  -0.50 &   2.2 \\
GJ 3494  &  4470 &  4.9 &  -0.25 &   3.3 \\
\end{longtable}

\tiny
\begin{center}
\begin{landscape}
\begin{longtable}{|lcccccccccc|}
  \caption[ ]{Stellar parameters (e.g. $T_{eff}$, $\pi$, M$_V$, R$_{\star}$) 
as well as the values of $v\sin i$ from Paper~XIV for our dM2 stars. We 
computed the projected rotational period $P/\sin i$ as well as [M/H] derived 
from the [M/H]/radius correlation (Paper~XVII).} \\

\hline Star    & v    &(R-I)$_{c}$&  $T_{eff}$   &Spect.  &     $\pi$     &      M$_{v}$    &   R$_{\star}$    &$v\sin i$&   $P/\sin i$  & [M/H]   \\
        &(mag) & (mag)     &     (K)      & Type   &     (m'')     &      (mag)      &  (R$_{\odot}$)   & $km/s$  &    (days)     & (dex)   \\
        &      &           &              &        &               &                 &                  &         &               &         \\
\endfirsthead
\caption{continued.}\\

\hline Star    & v    &(R-I)$_{c}$&  $T_{eff}$   &Spect.  &     $\pi$     &      M$_{v}$    &   R$_{\star}$    &$v\sin i$&  $P/\sin i$   & [M/H]   \\
        &(mag) & (mag)     &     (K)      & Type   &     (m'')     &      (mag)      &  (R$_{\odot}$)   & $km/s$  &    (days)     & (dex)   \\
        &      &           &              &        &               &                 &                  &         &               &         \\
\endhead 
\hline
\endfoot

GJ 1010A&11.31 &   1.148   & 3638 $\pm$100& dM2    & 50.01$\pm$2.02&  9.805$\pm$0.106&  0.517 $\pm$0.065&    -    &      -        & -0.098  \\
GJ 1062 &13.015&   1.184   & 3602 $\pm$100&sdM2    &    33$\pm$3.6 & 10.608$\pm$0.245&  0.372 $\pm$0.076&   9.8   & 1.92-0.53+0.65& -0.584  \\
GJ 1114 &11.523&   1.148   & 3638 $\pm$100& dM2    & 54.53$\pm$3.60& 10.206$\pm$0.144&  0.430 $\pm$0.063&   3.8   & 5.73-1.50+2.07& -0.339  \\
GJ 1264 & 9.641&   1.112   & 3674 $\pm$100& PMS M2e& 61.29$\pm$1.54&  8.578$\pm$0.059&  0.870 $\pm$0.088&   6.46  & 6.82-1.51+2.06& +0.378  \\
GJ 2085 &11.185&   1.058   & 3724 $\pm$100& dM2    & 46.81$\pm$2.39&  9.537$\pm$0.113&  0.525 $\pm$0.067&   2.46  &10.80-3.23+5.31& +0.000  \\
GJ 3084 &10.791&   1.064   & 3718 $\pm$100& dM2    & 48.83$\pm$10.7&  9.234$\pm$0.435&  0.608 $\pm$0.187&   3.16  & 9.74-4.07+5.98& +0.023  \\
GJ 3098 &11.155&   1.088   & 3696 $\pm$100& dM2    & 55.98$\pm$1.91&  9.895$\pm$0.078&  0.461 $\pm$0.051&   2.60  & 8.97-2.49+3.98& -0.200  \\
GJ 3207 & 11.51&   1.160   & 3626 $\pm$100& dM2    &    57$\pm$14  & 10.289$\pm$0.497&  0.420 $\pm$0.147&    -    &      -        & -0.346  \\
\hline 
\end{longtable}
\end{landscape}
\end{center}

\tiny
\begin{center}
\begin{landscape}
\begin{longtable}{|lccccccccccc|} 
  \caption[ ]{Stellar parameters (e.g. $T_{eff}$, $\pi$, M$_V$, R$_{\star}$) 
as well as the values of $v\sin i$ from HM and [M/H] compiled from the 
literature for our targets in our list of dM3 stars. We computed the projected 
rotational period $P/\sin i$ as well as [M/H] derived from the [M/H]/radius 
correlation.} \\

\hline Star    & v    &(R-I)$_{c}$&     Teff      &Spect.  &     $\pi$     &       Mv       &    R$_{\star}$   &$v\sin i$&   $P/\sin i$  & [M/H] & [M/H]\footnotemark[1] \\
        &(mag) & (mag)     &     (K)       & Type   &     (m'')     &     (mag)      &   (R$_{\odot}$)  &($km/s$) &    (days)     & (dex) & (dex)                 \\
        &      &           &               &        &               &                &                  &         &               &       &             \\
\endfirsthead
\caption{continued.}\\

\hline Star    & v    &(R-I)$_{c}$&     Teff      &Spect.  &     $\pi$     &       Mv       &    R$_{\star}$   &$v\sin i$&  $P/\sin i$    & [M/H] & [M/H]\footnotemark[1] \\
        &(mag) & (mag)     &     (K)       & Type   &     (m'')     &     (mag)      &   (R$_{\odot}$)  &($km/s$) &     (days)     & (dex) & (dex)                 \\
        &      &           &               &        &               &                &                  &         &                &       &             \\
\endhead 
\hline
\endfoot

GJ 1046 &11.595&   1.292   & 3506 $\pm$100 & dM3    & 71.06$\pm$3.23& 10.85$\pm$0.101& 0.385$\pm$0.050  &  2.63   &  7.41-1.1+1.2  &   -   &-0.078 \\
GJ 1050 &11.730&   1.322   & 3484 $\pm$100 & dM3    &    66$\pm$13  & 10.83$\pm$0.435& 0.404$\pm$0.128  &  0.59   & 34.4-18+310    & -1.20 &-0.035 \\
GJ 1054B&13.095&   1.363   & 3456 $\pm$100 & dM3e   & 52.08$\pm$2.13& 11.68$\pm$0.091& 0.288$\pm$0.036  & 32.13   & 0.45-0.07+0.08 & -0.09 &-0.323 \\ 
GJ 1077AB&11.91&   1.38    & 3444 $\pm$100 & dM3    &  77.5$\pm$11  & 12.11$\pm$0.308& 0.243$\pm$0.060  &   -     &     -          & -0.64 &-0.499 \\
GJ 1080 &12.81 &   1.344   & 3469 $\pm$100 & dM3    &    50$\pm$3   & 11.30$\pm$0.150& 0.334$\pm$0.052  &   -     &     -          & -0.44 &-0.187 \\
GJ 1097 &11.456&   1.400   & 3431 $\pm$100 & dM3    & 81.38$\pm$2.49& 11.01$\pm$0.068& 0.415$\pm$0.048  & $<$0.5  &  $>$41.91      & -0.05 &-0.033 \\
GJ 1110AB&13.08&   1.324   & 3482 $\pm$100 & dM3    &  45.2$\pm$3.2 & 12.11$\pm$0.168& 0.225$\pm$0.037  &   -     &     -          & -0.71 &-0.579 \\
GJ 1125 &11.710&   1.125   & 3661 $\pm$100 &sdM3    &103.46$\pm$3.94& 11.78$\pm$0.085& 0.202$\pm$0.023  &  0.99  & 10.3-2.9+8.3   & -0.16 &-0.691 \\
\hline 
\end{longtable}
\footnotetext[1]{Metallicity from the radius-metallicity relation}
\end{landscape}
\end{center}

\tiny
\begin{center}
\begin{landscape}
\begin{longtable}{|lccccccccccc|} 
  \caption[ ]{Stellar parameters (e.g. $T_{eff}$, $\pi$, M$_V$, R$_{\star}$) 
as well as the values of $v\sin i$ and [M/H] compiled from the literature for 
our targets in our list of dM4 stars. We computed the projected rotational 
period $P/\sin i$ as well as [M/H] derived from the [M/H]/radius correlation.} \\

\hline Star    & v   &(R-I)$_{c}$& $T_{eff}$ & Spect. &     $\pi$     &       Mv       &  R$_{\star}$   &$v\sin i$& $P/\sin i$  & [M/H] & [M/H]\footnotemark[1] \\
        &(mag) & (mag) &     (K)      & Type   &     (m'')     &     (mag)      & (R$_{\odot}$)  &$(km/s)$ &   (days)    & (dex) & (dex) \\
        &      &       &              &        &               &                &                &         &             &       &       \\
\endfirsthead
\caption{continued.}\\

\hline Star    & v   &(R-I)$_{c}$& $T_{eff}$ & Spect. &     $\pi$     &       Mv       & R$_{\star}$    &$v\sin i$& $P/\sin i$  & [M/H] & [M/H]\footnotemark[1] \\
        &(mag) & (mag) &     (K)      & Type   &     (m'')     &     (mag)      & (R$_{\odot}$)  &$(km/s)$ &  (days)     & (dex) & (dex) \\
        &      &       &              &        &               &                &                &      &                &       &       \\
\endhead 
\hline
\endfoot

GJ 1001A&12.84 & 1.556 & 3297 $\pm$100 & dM4    & 76.86$\pm$3.97& 12.27$\pm$0.132& 0.291$\pm$0.044& 1.15 &12.81-5.66+18.02&   -   &+0.047 \\
GJ 1005AB&11.483&1.535 & 3315 $\pm$100 & dM4    &200.53$\pm$9.41& 12.99$\pm$0.104& 0.202$\pm$0.028& 3.00 & 3.41-0.96+1.44 & -0.47 &-0.210 \\
GJ 1010B&14.10 & 1.535 & 3315 $\pm$100 & dM4    &  63.7$\pm$5.3 & 13.12$\pm$0.201& 0.190$\pm$0.036&  -   &     -          &  0.01 &-0.249 \\
GJ 1013 &12.75 & 1.548 & 3304 $\pm$100 & dM4    &  62.3$\pm$4.2 & 11.72$\pm$0.167& 0.370$\pm$0.063&  -   &     -          & -0.07 &+0.064 \\
GJ 1034 &15.05 & 1.613 & 3250 $\pm$100 & dM4    &  48.1$\pm$4.5 & 13.46$\pm$0.224& 0.182$\pm$0.037& 4.50 & 2.05-0.61+0.79 & -0.38 &-0.275 \\
GJ 1045 &14.47 & 1.672 & 3211 $\pm$100 & dM4    &  48.8$\pm$3.4 & 12.91$\pm$0.172& 0.258$\pm$0.047&  -   &     -          &  0.30 &-0.018 \\
GJ 1065 &12.79 & 1.535 & 3315 $\pm$100 & dM4    &   104$\pm$3   & 12.88$\pm$0.083& 0.213$\pm$0.027& 4.00 & 2.70-0.65+0.88 &  0.20 &-0.172 \\
GJ 1068 &13.58 & 1.643 & 3231 $\pm$100 & dM4    &    80$\pm$22  & 13.10$\pm$0.633& 0.226$\pm$0.104&  -   &     -          &  0.06 &-0.126 \\
\hline  
\end{longtable}
\footnotetext[1]{Metalicity from the radius-metallicity relation}
\end{landscape}
\end{center}

\newpage
\begin{center}
\begin{table}
\caption[ ]{Mean parameters of our stellar samples. Notably, we give the mean 
effective temperature and the convective overturning time $\tau_{c}$ according 
to Noyes et al. (1984) and Spada et al. (2013).}
\begin{tabular}{|lccccccc|}

\hline

Spectral & (R-I)$_{c}$ &(R-I)$_{c}$&T$_{eff}$&R$_{\star}$ & (B-V)&$\tau_{c}$\footnotemark[1]&$\tau_{c}$\footnotemark[2]\\
Type     & domain (mag)&  (mag)    &  (K)   &(R$_{\odot}$)& (mag)&    (days)   &    (days) \\
         &             &           &        &             &      &             &           \\
dK4      &[0.560;0.680]&   0.617   &  4600  &   0.815     & 1.151&   24.16     &  51.3     \\
dK6      &[0.684:0.816]&   0.738   &  4226  &   0.6469    & 1.340&   25.68     &  76.5     \\
dM2      &[1.064:1.196]&   1.128   &  3658  &   0.5082    & 1.511&   27.14     &  119      \\
dM3      &[1.284:1.416]&   1.354   &  3462  &   0.4090    & 1.548&   27.46     &  175      \\
dM4      &[1.500:1.700]&   1.532   &  3317  &   0.3008    & 1.581&   27.75     &  454      \\
         &             &           &        &             &      &             &           \\
\hline 
\end{tabular}
\footnotetext[1]{From Noyes et al. 1984.}
\footnotetext[2]{From Spada et al. 2013.}
\end{table}
\end{center}
\normalsize

\tiny
\begin{longtable}{|lccccccc|}
\caption[ ]{Values of $v\sin i$ which we have derived for the stars in our 
late-K stellar sample. Also listed: FWHM of the cross-correlation profile, and 
$P/\sin i$.} \\

\hline Star    &     FWHM      &       FWHM    &V$_{rad}$&    $v\sin i$    &    $v\sin i$     &$v\sin i$& $P/\sin i$\\
        &(\AA$\pm 3\sigma$)&(\AA$\pm 3\sigma$)&km s$^{-1}$&km s$^{-1}\pm 3\sigma$&km s$^{-1}\pm 3\sigma$&km s$^{-1}$& (days$\pm 3\sigma$)  \\
        &     HARPS     &    SOPHIE     &         &      HARPS      &        SOPHIE    & Others  &        \\
\endfirsthead
\caption{continued.}\\

\hline Star    &     FWHM      &       FWHM    &V$_{rad}$&    $v\sin i$    &    $v\sin i$     &$v\sin i$& P/sini\\
        &(\AA$\pm 3\sigma$)&(\AA$\pm 3\sigma$)&km s$^{-1}$&km s$^{-1}\pm 3\sigma$&km s$^{-1}\pm 3\sigma$&km s$^{-1}$& (days$\pm 3\sigma$)  \\
        &     HARPS     &    SOPHIE     &         &      HARPS      &        SOPHIE    & Others  &        \\
\endhead 
\hline
\endfoot

GJ 1056 & 0.1327 0.0028 &     -         & +5.05   &  1.82-0.38+0.32 &         -        &    -    & 19.1+7.2-4.3   \\
GJ 1066 & 0.1334 0.0028 &     -         & +26.08  &  1.91-0.37+0.29 &         -        &    -    & 17.7+6.0-3.5   \\
GJ 1067 &      -        & 0.1672 0.0034 & -10.10  &        -        &  2.08-0.62+0.45  &    -    & 16.2+9.3-4.3   \\
GJ 1177A&       -       &     -         &    -    &         -       &         -        &   6.1   & 4.78+1.5-1.1   \\
GJ 1213 &       -       &     -         &    -    &         -       &         -        &  0.0    &   $>$31.6      \\
GJ 1267 & 0.1366 0.0028 &     -         & -5.18   &  2.24-0.28+0.27 &         -        &    -    & 13.7+3.4-2.6   \\
GJ 1279 & 0.1347 0.0028 &     -         & +4.81   &  2.06-0.35+0.27 &         -        &  1.0    & 17.1+5.2-3.3   \\
GJ 3072 & 0.1277 0.0028 &     -         & +9.22   &  1.10-0.53+0.42 &         -        &    -    & 26.3+31-9.6    \\
\hline 

\end{longtable}
\footnotetext[1]{Rutten 1987}
\normalsize

\end{document}